\documentclass[11pt]{article} 
\usepackage[utf8]{inputenc}
\usepackage{multirow}
\usepackage{amsfonts}
\usepackage{amsmath}
\usepackage{amssymb}
\usepackage{amsthm}
\usepackage{caption}
\usepackage[dvipsnames]{xcolor}
    \definecolor{darkgreen}{rgb}{0,0.5,0}
    \definecolor{darkblue}{rgb}{0,0,0.6}
    \definecolor{purple}{rgb}{0.4,.2,0.7}
\usepackage[margin = 2.5cm]{geometry}
\usepackage{graphicx}
\usepackage[hyperfootnotes = false, colorlinks = true, linkcolor = darkblue, citecolor = purple]{hyperref}
\usepackage{subcaption}

\newcommand{\be}{\begin{equation}}
\newcommand{\ee}{\end{equation}}
\newcommand{\bea}{\begin{eqnarray}}
\newcommand{\eea}{\end{eqnarray}}
\def\la{\label}
\def\nref#1{(\ref{#1})}
\def\half{{1 \over 2 }}

\def\bG{ {{\bar G}_\phi} }

\newcommand{\lp}{\left (}
\newcommand{\rp}{\right )}
\newcommand{\lb}{\left [}
\newcommand{\rb}{\right ]}
\usepackage{tensor}

\newcommand{\pder}[2]{\frac{\partial#1}{\partial#2}} 

\newcommand{\cJ}{\mathcal{J}}
\DeclareMathOperator{\sgn}{sgn}
\DeclareMathOperator{\erfi}{erfi}
\newcommand{\mean}{\overline}
\newcommand{\rsp}{q} 
\newcommand{\ssc}{X} 
\DeclareMathOperator{\pfaff}{Pf}

\begin{document}

\thispagestyle{empty}
\begin{center}
    ~\vspace{5mm}

  {\LARGE \bf {A supersymmetric SYK model with a curious  low energy behavior \\}}

   \vspace{0.5in}
     
   {\bf    Anna Biggs$^1$, Juan Maldacena$^2$ and Vladimir Narovlansky$^1$
   }

    \vspace{0.5in}

  $^1$
  Jadwin Hall, Princeton University,  Princeton, NJ 08540, USA 
   \\
   ~
   \\
  $^2$
  Institute for Advanced Study,  Princeton, NJ 08540, USA

    \vspace{0.5in}

    \vspace{0.5in}

\end{center}

\vspace{0.5in}

\begin{abstract}
 
We consider ${\cal N}=2,4$ supersymmetric SYK models that have a peculiar low energy behavior, with the entropy going like $S=S_0 + ({\rm constant}) T^a$,  where $a \not =1$. The large $N$ equations for these models are a generalization of equations that have been previously studied as     an unjustified truncation of the planar diagrams describing the BFSS matrix quantum mechanics or other related  matrix models.  Here we reanalyze these equations  in order to better understand the low energy physics of these models. We find that the scalar fields develop large expectation values which explore the low energy valleys in the potential. The low energy physics is dominated by quadratic fluctuations around these values. These models were previously conjectured to have a spin glass phase. We did not find any evidence for this phase by using the usual diagnostics, such as searching for replica symmetry breaking solutions.  
  
 \end{abstract}
 
\vspace{1in}

\pagebreak

\setcounter{tocdepth}{3}
{\hypersetup{linkcolor=black}\tableofcontents}
  
    
  \section{Introduction and Motivation}

  We study ${\cal N}=2,4$ supersymmetric SYK-like models with dynamical bosons and fermions in 0+1 dimensions. 
  In the past, these models have appeared in three somewhat unrelated lines of development.
  
  The first motivation comes from the  BFSS matrix model \cite{Banks:1996vh}.  As discussed in  \cite{Itzhaki:1998dd}, this model can be studied in an energy regime where the entropy scales as $S \propto T^{9/5}$ and it can be viewed as a very simple example of the duality between a quantum mechanical system and a gravity theory described by the Einstein  equations. 
   This prompted  \cite{Kabat:1999hp,Kabat:2000zv,Kabat:2001ve} to try to reproduce this temperature  scaling  from the matrix model. In the process,  they produced various truncations of the model and they summed a subset of the planar diagrams. This produced results where $S = S_0 + \# T^{6/5}$, and even a power law of nearly $S = S_0 + \# T^{9/5}$,  depending on the precise truncation, see also \cite{Lin:2013jra}.\footnote{See eq. (18) of \cite{Kabat:1999hp}, eq. (47) of \cite{Kabat:2001ve}, and eq. (3.27) of \cite{Lin:2013jra}.}
   Here we point out that some of the equations analyzed in these papers are actually the large $N$ equations of ${\cal N}=2,4$ supersymmetric SYK models. In other words, those equations came from an unjustified truncation for the models discussed in those papers, but they are the right equations to describe particular cases of the ${\cal N}=2,4$ models discussed in this paper. Therefore we can use their analysis of the equations to discuss the models in this paper. 
   
   The second motivation arises as follows. The ${\cal N}=4$ version of the model was written before by Anninos, Anous and Denef as a model for black holes and black hole fragmentation \cite{Anninos:2016szt}. They suggested that the model could have a spin glass phase, similar to that observed in purely bosonic models \cite{Cugliandolo_2001,Anous:2021eqj}.  We will analyze this question further and we will answer it in the negative, at least for the specific model we consider.   

  Finally, the higher dimensional version of these models was studied in \cite{Murugan:2017eto} where they noticed that the low energy physics is a conformal field theory. In other words, we are studying the 0+1 dimensional version of the SYK-like models introduced in \cite{Murugan:2017eto}, see also \cite{Popov:2019nja} for a tensor model version.  
      
        In this paper we explore the physics of these models, putting together the various threads of developments that we mentioned above. 
       The low energy thermodynamics is rather peculiar. It is neither trivially gapped nor conformal as in the usual SYK model.  
    The low energy expansion of the large $N$ equations was previously studied in  \cite{Lin:2013jra}.   We will present it in a slightly simpler way through an expansion of the large $N$ effective action. Besides simplifying the derivation,  this suggests a physical interpretation  for and the reason behind the fractional powers of the temperature. The physical picture involves the scalar fields exploring shallow valleys in the potential landscape. Fluctuations around these shallow valleys become very slightly coupled harmonic oscillators. This probably implies that these models are too simple to model interesting black holes.

      

    This paper is organized as follows. 
    In section 2, we define the models and discuss some simple properties. In section 3, we write the large $N$ equations and we analyze some aspects of the solution. In section 4, we discuss features of the numerical solutions of the equations. In section 5, we perform the low temperature expansion of the large $N$ equations. In section 6, we look for a spin glass phase as well as a dynamical phase by looking for 1 step replica symmetry breaking solutions. We argue that there are no solutions in the physical range of parameters. In section 7, we offer a physical interpretation of the low energy physics, as well as a comparison to some related models.

    \section{Definition of the models and simple properties }

  The models we study come in two types, an ${\cal N}=2$ version and an ${\cal N}=4$ version. They are very similar for the physics we will be interested in. Let us introduce them in turn. 
  
  \subsection{ ${\cal N}=2$ supersymmetric SYK  model with dynamical bosons and fermions } 
  
  This model has two supersymmetries. We write both the superspace Lagrangian as well as the component one.   
 
 To use superspace, we  introduce anticommuting $\theta ^{\alpha } $, $\alpha =1,2$. The R-symmetry is $SO(2)$ so there is no good distinction between upper and lower indices, but we will use a convention where we lower and raise indices with the $\epsilon $ tensor. As a convention we can take for instance $\epsilon _{12} =\epsilon ^{21} =1$. In addition to $\delta _{\alpha } ^{\beta } $, we also have the invariant tensor $\delta _{\alpha \beta } $.  We define $\theta ^2=\frac{i}{2} \theta ^{\alpha } \theta _{\alpha } $. The super-derivative is
\begin{equation}
D_{\alpha } = \pder{}{\theta ^{\alpha } } + \theta ^{\alpha } \pder{}{\tau } 
\end{equation}
and we will work in Euclidean time. The superfield is
\begin{equation} \label{eq:superfield}
\Phi =\phi +i\theta ^{\alpha } \psi_{\alpha } +i\theta ^2F .
\end{equation}

We take $N$ such superfields, so we have a dynamical boson $\phi^i$, two Majorana fermions $\psi^i_\alpha$ and an auxiliary field $F^i$ for each value of $i=1,\dots, N$.
The action is
\begin{equation}
\begin{split} \la{LagTwoC}
S&=\int d\tau d^2\theta \, \left[ - \frac{i}{4}  \epsilon ^{\alpha \beta } D_{\alpha } \Phi^i D_{\beta } \Phi^i + C_{ijk} \Phi ^i \Phi ^j \Phi ^k \right] =\\
& = \int d\tau \left[ \frac{1}{2} \psi^i_{\alpha } \dot \psi^i_{\alpha } +\frac{1}{2}(F^i)^2 +\frac{1}{2} (\dot \phi^i) ^2 +3i C_{ijk} \left( F^i \phi ^j \phi ^k + \epsilon ^{\alpha \beta } \psi^i_{\alpha } \psi^j_{\beta } \phi ^k \right) \right] 
\end{split}
\end{equation}
  where the coefficients $C_{ijk}$ are real and we define them to be symmetric in the indices. 
We can integrate out the fields $F^i$ to obtain a bosonic potential which has the form $ \sum_i [\partial_i W(\phi) ]^2$, where $W(\phi) = C_{ijk} \phi^i \phi^j \phi^k $. 
We can also consider a generalization where the superpotential is changed   
\be 
 C_{ijk} \phi^i \phi^j \phi^k \to  C_{i_1 i_2\cdots i_p} \phi^{i_1} \phi^{i_2} \cdots  \phi^{i_p} 
 \ee
 where $p$ is any integer. 
 
 The model has a continuous $SO(2)$ R-symmetry\footnote{``R-symmetry'' means a global symmetry that does not commute with supersymmetry and which rotates the two supercharges.} that rotates the two fermions into each other. 
    
    In addition, for odd $p$,  we also have a $\mathbb{Z}_2$ R-symmetry flipping the sign of one of the fermions and the sign of the bosons, 
    $\psi_1^i \to \psi^i_1 , ~ \psi_2^i \to -\psi_2^i ,~ \phi^i \to - \phi^i$.
    For even $p$ there is a $\mathbb{Z}_2$ symmetry under which all fields $\phi^i$, $\psi^i_{\alpha}$ and $F^i$ are odd.
    
 For $p$ odd we have time reversal symmetry under which $\phi ^i \to -\phi ^i$ and $F^i \to -F^i$, leaving the fermions invariant. This is similar to the discussion in \cite{Gaiotto:2018yjh}, where a time reversal was used  in a three dimensional context with the same amount of supersymmetry.
 For $p$ even we have time reversal symmetry taking $\phi^i \to -\phi ^i$ and $\psi _2 \to -\psi _2$; it does not commute with supersymmetry.

    \subsection{  ${\cal N}=4$ supersymmetric SYK  model with dynamical bosons and fermions }
    
    This model was first discussed by Anninos, Anous and Denef \cite{Anninos:2016szt}\footnote{They actually discussed a slighly different version with $3 N$ superfields and extra $U(1)$ global symmetries.}. 
    It can be written introducing two complex anticommuting variables $\theta^{\alpha }$ and their complex conjugates $\bar \theta_\alpha$  which can be viewed as arising from the dimensional reduction of four-dimensional ${\cal N}=1$ superspace to one dimension.\footnote{When reducing from four dimensions, an $SO(3) \subset SO(4)$ that keeps the time invariant is singled out. Then   $\sigma ^{\mu =0} _{\alpha \dot \alpha } $ becomes an invariant tensor.}
   We define 
   \be 
   D_{\alpha } = \pder{}{\theta ^{\alpha } } + \bar \theta _{  \alpha } \pder{}{\tau } ~,~~~~~~~ \bar D^\alpha   = \pder{}{\bar \theta _{  \alpha } } +   \theta ^{  \alpha } \pder{}{\tau }
   \ee 
   so that $\{D_{\alpha } ,\bar D^{\beta } \}=2\delta _{\alpha } ^{\beta } \partial _{\tau } $.
   We have complex chiral superfields obeying $\bar D^{\alpha } \Phi^i=0$ and 
   $D_{\alpha } \bar \Phi^i =0$. 
   They can be expanded as 
   \begin{equation}
   \begin{aligned}
      \Phi^i( \theta^\alpha, y) &= \phi^i(y) + \sqrt{2} \theta^\alpha \psi_{\alpha}^i(y) + \theta^2 F^i(y) ,\qquad && y = \tau + \theta^{\alpha }  \bar \theta_{\alpha }  ,\\
   \bar \Phi ^i(\bar \theta _{\alpha } ,\bar y) &= \bar \phi ^i(\bar y)-\sqrt{2}\bar \theta _{\alpha } \bar \psi ^{i\alpha }(\bar y) +\bar \theta ^2\bar F^i(\bar y),\qquad && \bar y = \tau -\theta ^{\alpha } \bar \theta _{\alpha } 
   \end{aligned}
   \end{equation}
   where in $\mathcal{N} =4$ discussions we use the different convention $\theta ^2=\theta ^{\alpha } \theta _{\alpha } $ and $\bar \theta ^2=\bar \theta ^{\alpha } \bar \theta _{\alpha } $.
   
   The action is   given by 
   \bea 
   S &=& \int d \tau d^4 \theta \, \bar \Phi^i \Phi^i + \left( \int d\tau d^2 \theta \, W(\Phi) + c.c \right)
   ~,~~~~~~~W = {\cal C}_{ijk} \Phi^i \Phi^j \Phi^k 
    \cr 
    &=& \int d \tau  \left[  \dot{ \bar \phi}^i \dot \phi^i + \bar \psi^{\alpha i } \dot \psi_{\alpha }^i + \bar F^i F^i + 3{\cal C}_{ijk} \left( F^i \phi^j \phi^k + \epsilon^{\alpha \beta } \psi^i_\alpha \psi_\beta^j \phi^k \right) + c.c. \right] 
    \eea
   where the coefficients ${\cal C}_{ijk}$ are complex and symmetric.     We have a complex dynamical boson, two complex fermions and a complex auxiliary field for each index $i=1, \cdots, N$. 
    Again, we can integrate out the auxiliary field to obtain a bosonic potential 
    $V \propto \sum_i |\partial_i W|^2 $. 

    This action has an $SU(2) \times U(1) $ R symmetry.  In terms of the dimensional reduction from four dimensions, the $SU(2)$ is the part of the $SO(4)$ that remains after we single out the time direction, and the $U(1)$ is the R-symmetry under which $[\phi ]=2/3$, $[\psi ]=-1/3$ and $[F]=-4/3$ for $p=3$. The same generalization to general $p$ exists in the $\mathcal{N} =4$ version. So there is less distinction between even or odd $p$.

    \subsection{The Witten index and supersymmetric ground states}
    
    In supersymmetric quantum mechanics, the energy is positive semidefinite. States with zero energy are annihilated by the supercharges and preserve supersymmetry.  It is interesting to understand whether we have a large number of such states since they might appear as a contribution to the ground state entropy. 
        
    A useful sufficient condition for the existence of zero energy states is a nonzero value of the Witten index \cite{Witten:1982df}, or 
    \be \la{WitInd}
     {\cal I }_W = Tr[ (-1)^F e^{ -\beta H } ] 
     \ee 
     where $F$ is the fermion number operator.  
     
     We provide a brief review for readers unfamiliar with the Witten index. ${\cal I }_W $ is similar to the partition function but differs by the insertion of the fermion parity operator $(-1)^{F}$, which takes eigenvalues $+1$ on bosonic states and $-1$ on fermionic states. In supersymmetric theories, all non-zero energy states are degenerate. Specifically, for every bosonic state with non-zero energy there is a fermionic state with equal energy. This is argued as follows, the supersymmetry algebra $\{ Q^\dagger, Q \} = H $, $[H, Q]= [H,Q^\dagger] =0$,  implies that,  in the subspace of states with energy $E$,  $Q$ and $Q^\dagger$ act like fermionic creation and annihilation operators, giving rise to a pair of states, both with the same energy, but opposite fermion number.    This means that contributions to ${\cal I }_W $ from non-zero energy states cancel out, leaving only contributions from the ground states. This further implies that the result is independent of $\beta$, since all states with non-zero energies are paired and therefore cancel in \nref{WitInd}.  In summary, the Witten index, computed at any temperature, counts the difference in the number of bosonic and fermionic ground states, and ${\cal I }_W \neq 0$ implies that there are some ground states.  
     Furthermore, we can also see that  the Witten index is independent of subleading deformations of the superpotential. The reason is that zero energy states can only appear and disappear by turning into a boson and fermion pair of states with non-zero energy. While this changes the total number of zero energy states, it does not change the difference between the number of bosonic and fermionic zero energy states.

     \subsubsection{The Witten index for the ${\cal N}=4$ case} 
     
     The Witten index is easiest to compute in the ${\cal N}=4$ case \cite{Anninos:2016szt}. 
     It is convenient to introduce a subleading deformation of the superpotential. The simplest subleading deformation is to add a linear term $\tilde W = {\cal C}_{i_1 \cdots i_p}  \Phi^{i_1}\cdots \Phi ^{i_p}  + v_i \Phi^i $. The potential energy has a classical zero energy minimum when the superpotential has a critical point.  With the linear deformation, the condition for such minima becomes
     \be 
      \partial_i \tilde W = p {\cal C}_{i i_2 \cdots i_p} \phi^{i_{2} } \cdots \phi^{i_p} + v_i =0   .
      \ee 
    We expect that for generic values of   ${\cal C}$  and $v$ we only have isolated solutions. Since we have $N$ equations, each of order $(p-1)$ we expect 
    $(p-1)^N$ solutions \cite{Garcia_Li}. Around each solution we can approximate the superpotential as quadratic. For a quadratic potential, we have exactly one zero energy state which is bosonic.\footnote{Around each minimum we have a fermion mass term $  i m ( \psi_1 \psi_2 + \bar \psi^1 \bar \psi^2 ) =   i m  ( c_1 c_2 + c_1^\dagger c_2^\dagger)$ where $m \propto \partial^{2}  W$. This term gives rise to four states. The lowest energy state, which is a combination of the state annihilated by the $c_i$ and the state annihilated by the $c_i^\dagger$,  is the only one that can cancel the ground state energy of the complex boson and give rise to a zero energy state \cite{Witten:1982df}.} Note that by making $v$ very large we can make sure that this perturbative computation around each critical point can be trusted. This means that we have 
    \be 
    {\cal I}_W = (p-1)^N ~,~~~~~~~~{\rm for }  ~~{\cal N}=4 .
    \ee 
    So the conclusion is that we have a large number of zero energy states. Therefore we expect a non-zero entropy of at least $S_0 = N \log(p-1)$ for this case. 
    
    \subsubsection{The Witten index for the ${\cal N}=2$ case}
    
    For the ${\cal N}=2$ case we use a similar logic where we add a linear term to the superpotential, see 
    \cite{Ghim:2019rol} for this type of index computations.    Again we expect $(p-1)^N$ solutions. However there are some new features. First the solutions could be complex, but our fields are real, so we are only interested in real solutions. The second issue is that even for real solutions the contribution to the index can be positive or negative depending on the sign of the fermion mass term. 
    Namely, if we had a single field with a quadratic mass term, the fermion mass term would be $ i m \psi_1 \psi_2$. So the lowest energy state could be a boson or a fermion depending on the sign of $m$. For our problem,  around  each solution of $\partial_i \tilde W =0$, we have a quadratic theory with a mass matrix $\partial_i \partial_j W$ evaluated at the solution. Whether the state is a fermion or a boson is determined by the sign of the determinant of this matrix, $\det (\partial_i \partial_j W)$. 
    
    In the particular case of odd $p$   there is a simplification. If $\phi_*^i$ is a solution, then $ - \phi_*^i$ is a solution since $\tilde W$ would be odd in this case. Furthermore, this implies that the fermion mass matrix changes sign as we flip the sign of $\phi_*^i$.    If,  in addition, $N$ is odd, its determinant also  changes sign. This implies that the contribution to the index vanishes. 
    
    Then we conclude that 
    \be 
    {\cal I }_W = 0 ~,~~~~~~{\rm for } ~~{\cal N}=2,~~~~p ~~{\rm odd}, ~~{\rm and }~~ N ~~{\rm odd } .
    \ee 
    
    We can further explicitly check that for the case $N=1$ and $p$ odd, there are no normalizable zero energy solutions. So for this case there are no zero energy states. We expect that the same is true for larger odd values of $N$. 
    (For $N=1$ and even $p$ the index is $\pm 1$). 
            
    As a side comment, we note that we cannot hope to prove that ${\cal I}_W =0$ for even $N$ since the ${\cal N}=4$ case can be viewed as a special case of an ${\cal N}=2$ theory with $2 N $ fields.\footnote{We thank Donwook Ghim for this comment.}
Indeed, we have numerically checked for small even values of $N$ (up to $N=12$) and various samplings from the Gaussian distribution for $C_{i_1 \cdots i_p}$ that the Witten index can be nonzero,  but  small relative to the number of real solutions.\footnote{In particular, for $N$ even and $p$ odd, $\mathcal{I}_{W}$ is an even integer. This is again due to the fact that if $\phi_*^i$ is a solution, then $ - \phi_*^i$ is also a solution, and the sign of $\det (\partial_i \partial_j W)$ is equal on the two solutions.} 

\

    \section{The large $N$ equations }\label{lgNeq}

     In this section,  we write the equations that determine the large $N$ solution of the model. 
 
The derivation of these equations involves averaging over the couplings $C$. Here we average the partition function, which is also called the annealed calculation. In principle, the correct way to treat the disorder is to do the quenched calculation, or average the free energy, which can be done by introducing replicas and taking the $n\to 0 $ limit. 
We will discuss replicas in section \ref{Replicas}, and it turns out that we will not find other interesting solutions when they are included. Therefore we will ignore the replicas, for now. 
   
  We start from the original path integral and we average over disorder by introducing a Gaussian integral over the couplings. We denote such an averaging by $\mean{(\cdots)}$. In the $\mathcal{N} = 2$ theory, we take the variance of each coefficient to be  
  \be \la{Cave}
\mean  { C_{i_1, \cdots , i_p}^2 } = { J^p \over   p! \, p  N^{p-1} }
\ee 
This is the variance when all the indices are different.\footnote{The factor of $p!$ is included because $C$ is a symmetric tensor, and each index is summed from 1 to $N$.}  We have a choice of how we pick the variance when some of the indices are the same, but these cases are subleading in the large $N$ expansion and it will not be important for what we will do in this paper.\footnote{ A particular choice where the formulas look simpler is to choose the variances such that if the distinct indices in $C_{i_1 \cdots i_p} $ appear $a_1,a_2,\cdots $ times (that is, if among the $p$ indices we have $k$ distinct indices $j_1,\cdots ,j_k$ each appearing $a_1,\cdots ,a_k$ times) then $\mean{ C^2_{i_1 \cdots i_p} }= \frac{J^p}{pN^{p-1} } \binom{p}{a_1,a_2,\cdots } ^{-1} $. This is the convention we will use before we take the large $N$ limit.}  

The derivation is most compact when written in superspace. Coordinates in superspace can be denoted collectively by $\ssc=(\tau ,\theta ^{\alpha } )$.
We would like to go to bi-local fields in superspace so that $G(\ssc_1,\ssc_2)$ represents
$G(\ssc_1,\ssc_2) = \frac{1}{N} \sum _i \Phi ^i(\ssc_1) \Phi ^i(\ssc_2) $.

After integrating over the disorder,\footnote{In the following we denote the averaged partition function still by $Z$ for simplicity, hoping it will not cause confusion.} the partition function is given in terms of the original fields as
\begin{equation}
\begin{split}
Z &= \int {\cal D}\Phi \exp \Bigg[ \frac{i}{4} \int d\ssc \, \epsilon ^{\alpha \beta } D_{\alpha } \Phi ^i D_{\beta}  \Phi ^i +\frac{J^p}{2pN^{p-1}} \int d\ssc d\ssc' \, \Phi ^{i_1} \cdots \Phi ^{i_p}(\ssc) \Phi ^{i_1} \cdots \Phi ^{i_p}(\ssc') \Bigg] .
\end{split}
\end{equation}
We introduce bilocal superfields $G(\ssc,\ssc')$ and $\Sigma (\ssc,\ssc')$, including all superspace components, and insert the identity
\begin{equation}
1=\int {\cal D}G {\cal D}\Sigma \, \exp \Bigg[-\frac{N}{2} \int d\ssc d\ssc' \, \Sigma (\ssc,\ssc')\left( G(\ssc,\ssc')-\frac{1}{N} \Phi ^i(\ssc) \Phi ^i(\ssc')\right)  \Bigg]
\end{equation}
so that
\begin{equation}
\begin{split}
& Z = \int {\cal D}\Phi {\cal D}G{\cal D}\Sigma  \exp \Bigg[ \frac{i}{4} \int d\ssc \, \epsilon ^{\alpha \beta } D_{\alpha } \Phi ^i D_{\beta } \Phi ^i + \frac{J^p N}{2p} \int d\ssc d\ssc' \, G(\ssc,\ssc')^p - \\
& \quad - \frac{N}{2} \int d\ssc d\ssc' \, \Sigma (\ssc,\ssc')\left( G(\ssc,\ssc') - \frac{1}{N} \Phi ^i(\ssc) \Phi ^i(\ssc') \right)  \Bigg].
\end{split}
\end{equation}

The $G\Sigma $ equations of motion are then simple in superspace
\begin{equation}
\begin{split}
& \Sigma (\ssc,\ssc') = J^p G(\ssc,\ssc')^{p-1},\\
& D_{\ssc}^2 G(\ssc,\ssc')+2i \int d\ssc'' \, \Sigma (\ssc,\ssc'') G(\ssc'',\ssc') = 2\delta (\ssc-\ssc') .
\end{split}
\end{equation}
Above we have defined $D^2=\epsilon ^{\alpha \beta } D_{\alpha } D_{\beta } $ and the delta function in superspace is $\delta (\ssc-\ssc')=\delta (\tau -\tau ') (\theta ^1-\theta ^{1\prime}) (\theta ^2-\theta ^{2\prime})$.

In fact, below we will need the explicit expressions in terms of the component fields. When looking for a large $N$ solution, we can use $SO(2)$ and the $\mathbb{Z} _2$ R-symmetry discussed above for odd $p$, assuming they are not broken, in order to deduce that it is enough to consider $G_{\phi \phi } $, $G_{\psi \psi } $ which is the same for $\langle \psi _1\psi _1\rangle $ and $\langle \psi _2\psi _2\rangle $, and $G_{FF} $. In other words, we are setting to zero some correlators such as $G_{\phi F} \propto \langle \phi^i F^i\rangle $ or $\langle \psi^i \phi^i \rangle$, see  appendix \ref{sec:Phi_F} for further discussion. 
We are left with
\begin{equation}
G(\ssc,\ssc') = G_{\phi \phi } (\tau-\tau ' ) + \theta ^{\alpha } \theta ^{\prime \alpha  } G_{\psi \psi  } (\tau -\tau ') -\theta ^2\theta ^{\prime 2} G_{FF} (\tau -\tau ').
\end{equation}
The expectation values depend only on the time difference. The signs above are chosen according to the superfield notation \eqref{eq:superfield}. Correspondingly,
\begin{equation}
\Sigma (\ssc,\ssc') = -\Sigma _{FF} (\tau -\tau ')+\theta _{\alpha } \theta '_{\alpha  } \Sigma _{\psi \psi  } (\tau -\tau ')+\theta ^2\theta ^{\prime 2} \Sigma _{\phi \phi } (\tau -\tau ').
\end{equation}
In addition, for notational convenience, we will denote bilocal fields of the same field by writing the field only once, such as $G_{\phi } \equiv G_{\phi \phi } $.


We can then integrate out all the original fields and be left with an effective action in the large $N$ theory  
  \begin{equation}
\begin{split} \la{EAFirst}
& {\log Z \over N }   =   - \half \log \det[ - \delta(\tau -\tau')\partial_\tau^2 - \Sigma_\phi ] - \half \log \det[ \delta(\tau-\tau' )  - \Sigma_F]  + 2 \log \pfaff[  \delta(\tau -\tau') \partial_\tau - \Sigma_\psi ] \\ 
& \qquad \qquad  -  \int d\tau d\tau ' \Big(  \Sigma _{\psi} G_{\psi} + \half \Sigma _{\phi } G_{\phi }+ \half \Sigma _F G_F -\frac{p-1}{2} J^p G_{\psi} ^2 G_{\phi } ^{p-2}+\frac{J^p}{2} G_F G_{\phi } ^{p-1} \Big) \Bigg].
\end{split}
\end{equation}
 We will mainly consider the euclidean time finite temperature case. In that case, the functions depend only on the time differences and the equations of motion become   
\begin{equation} \label{eq:SD_eqs_Sigma}
\begin{split}
1 =  [{\omega ^2- \Sigma  _{\phi   }(\omega) } ]G_{\phi  } (\omega )= [{-i \omega- \Sigma  _{\psi   }(\omega) } ]G_{\psi  } (\omega )=[{1- \Sigma  _{F   }(\omega) } ]G_{F  } (\omega )
\end{split}
\end{equation}
where the   Matsubara frequencies are $\omega_n = { 2 \pi \over \beta } n $ for bosons and $\omega_{r} = { 2 \pi \over \beta } r $, $r = n + \half$ for fermions, with $n$ integer.\footnote{We use the convention $G(\omega )=\int_0^\beta d\tau \, e^{i \omega \tau } G(\tau )$ for the Fourier transform.} 
We also have the equations  
 \begin{equation} \label{eq:SD_eqs_G_general_q}
\begin{split}
 \Sigma _{\phi } &=-(p-1)J^p G_F G_{\phi }^{p-2}  +(p-1)(p-2)J^p G_{\psi }^2 G_{\phi } ^{p-3} ,\\
 \Sigma _{\psi } &= (p-1) J^p G_{\psi } G_{\phi} ^{p-2} ,\\
 \Sigma _{F} &= -J^p G_{\phi  } ^{p-1}
\end{split}
\end{equation}
 which are in $\tau$ space. 
 
Let us discuss some general features of these equations. 
First, note that the mean expectation value of the squares of the scalars is  
\be 
{ 1 \over N } \sum_{i=1}^N \langle {\phi^i}^2 \rangle = G_{\phi } (\tau=0) 
\ee 
Note that \nref{eq:SD_eqs_Sigma} constrains the small $\tau$ behavior of the functions to be of the form 
\be \la{ShortD}
G_{\phi } = -\half |\tau | + \cdots ~,~~~~~~ G_{\psi } = \half {\rm sgn}(\tau) + \cdots ~,~~~~~~G_F(\tau) = \delta(\tau) + \cdots 
\ee 
where the dots denote analytic terms at small $\tau$. In particular, for $G_\phi$ these analytic terms can be larger than the terms indicated explicitly.

Based on the equations of motion of the auxiliary fields $F^i$ we see that 
\be 
G_F  = { 1\over N } \langle F^i(0) F^i(\tau) \rangle = -{ 1 \over N } \langle \partial_i W(0) \partial_i W(\tau) \rangle  ~,~~~~{\rm for } ~~~ \tau \not =0 
\ee 
where the minus sign comes from the $i$ in \nref{LagTwoC}. This implies that $G_F$ is expected to be negative for $\tau\not =0$. For $\tau=0$ we have a delta function with a positive coefficient  \nref{ShortD}.

As in the SYK case \cite{Maldacena:2018lmt}, it is possible to write an expression for the energy in terms of these functions 
\be 
{ E \over N }  = \half \left[   \partial_\tau^2  G_\phi - G_F +  2\partial_\tau G_\psi \right]_{\tau = 0^+} 
\ee 
   where we take the limit  $\tau \to 0$ from the positive direction.\footnote{In particular, this expression does not include the $\delta(\tau)$ in $G_F$.} We can derive this similarly to \cite{Maldacena:2018lmt} by considering the RHS, and replacing $\partial _{\tau } \psi |_{\tau=0}$ by $[H,\psi ]$, and similarly for the other terms. This gives manifestly the expectation value of the Hamiltonian $H$. This way we see that this relation holds also at finite $N$, and for any value of the couplings even without averaging.

In the $\mathcal{N} =4$ model, we normalize the couplings such that
\begin{equation}
\mean{ |{\cal C}_{i_1 \cdots i_p} |^2} =\frac{J^p}{p! p N^{p-1} } .
\end{equation}
Using the $SU(2) \times U(1)$ R-symmetry, if it is unbroken, 
 we again need to consider only three correlation functions $G_{\phi }(\tau,\tau') =\frac{1}{N} \langle \bar \phi^i (\tau )\phi^i (\tau ')\rangle $, $\frac{1}{N} \langle \bar \psi ^{\alpha i} (\tau )\psi _{\beta }^i (\tau ')\rangle  = \delta ^{\alpha } _{\beta } G_{\psi } (\tau ,\tau ')$ and $G_F(\tau ,\tau ')=\frac{1}{N}\langle \bar F^i(\tau ) F^i(\tau ')\rangle $.
In the $G\Sigma $ derivation we therefore need to include only these if we are interested in the large $N$ solution. Since now the fields are complex, it is useful to also include $\bar G_{\phi } (\tau ,\tau ')=\frac{1}{N} \phi ^i(\tau )\bar \phi ^i(\tau ')$, $2\bar G_{\psi } (\tau ,\tau ')=\frac{1}{N} \psi _{\alpha }^i (\tau )\bar \psi ^{\alpha i} (\tau ')$ and $\bar G_F(\tau ,\tau ')=\frac{1}{N}F^i(\tau )\bar F^i(\tau ')$; these are not new independent fields but rather are related by $\bar G_{A} (\tau ,\tau ')=(-1)^{F_A} G_{A } (\tau ',\tau )$ where $F_A$ is the fermion number of the field $A$ and similarly for $\Sigma $. By repeating a similar derivation, we get the $G\Sigma $ action 
\begin{equation}
\begin{split}
& {\log Z \over N }   =   - \log \det[ - \partial_\tau^2 - \bar \Sigma_\phi ] - \log \det[ \delta(\tau )  - \bar \Sigma_F]  +2 \log \det[   \partial_\tau - \bar \Sigma_\psi ] \\ 
& \qquad \qquad  -  \int d\tau d\tau ' \Big(  2\Sigma _{\psi} \bar G_{\psi} + \Sigma _{\phi }\bar G_{\phi }+ \Sigma _F \bar G_F -(p-1) J^p G_{\psi} ^2 G_{\phi } ^{p-2}+J^p G_F G_{\phi } ^{p-1} \Big) \Bigg].
\end{split}
\end{equation}
The Schwinger-Dyson equations that follow from this action are \eqref{eq:SD_eqs_Sigma} and
\begin{equation}
\begin{split}
 \Sigma _{\phi } &=-(p-1)J^p \bar G_F \bar G_{\phi }^{p-2}  +(p-1)(p-2)J^p \bar G_{\psi }^2 \bar G_{\phi } ^{p-3} ,\\
 \Sigma _{\psi } &= (p-1) J^p \bar G_{\psi } \bar G_{\phi} ^{p-2} ,\\
 \Sigma _{F} &= -J^p \bar G_{\phi  } ^{p-1}.
\end{split}
\end{equation}
The ${\cal N}=4$ theory has charge conjugation symmetry
\begin{equation}
\phi  \to \bar \phi ,\quad \psi _{\alpha } \to i \bar \psi ^{\alpha } \qquad \left( \text{and }F \to - \bar F \right) ;
\end{equation}
This is a symmetry of the disorder averaged theory (since it takes ${\cal C} \to \bar {\cal C})$. We assume that this symmetry is unbroken. Then $\bar G_A(\tau,\tau' )=G_A(\tau,\tau' )$ (and similarly for $\Sigma $) and the equations of motion are the same as in ${\cal N} =2$ and the action is just twice that of ${\cal N} =2$.
 We note that in our conventions $\bar F$ is minus the complex conjugate of $F$ and so $G_F$ is negative.

   \subsection{High temperature solution } 
   \la{HighEn}
   
   The relevant dimensionless parameter for these equations is $\beta J$. When it is small, we are in the high temperature regime. In this regime the theory is closer to the free theory. We analyze this regime here.  At high temperature we expect $\phi$ to fluctuate a lot and therefore $G_\phi$ should be relatively large. In addition, with a small size for the circle $\beta$ we expect that $G_\phi$ will be approximately independent of $\tau$. So we approximate $G_\phi = \bar G_\phi$ where $\bar G_\phi$ is a constant independent of $\tau$. 
  
   From the last term in \nref{eq:SD_eqs_G_general_q} we get $\Sigma_F = - J^p \bar G^{p-1}_\phi $, which in Fourier space implies  
\be \la{GFan}
 \Sigma_{F }(\omega=0)  = - \beta J^p {\bar G}^{p-1} ~,~~~~~~~ G_{F }(\omega=0) = { 1 \over 1 -   \Sigma_{F}(\omega=0) } = { 1 \over 1 + \beta  J^{p} {\bG}^{ p-1}  } .
 \ee 
Using the equation for $\Sigma_\phi$  in \nref{eq:SD_eqs_G_general_q}, neglecting the term involving $G_\psi$ and approximating $G_\phi$ as a constant  we get 
 \be 
   \Sigma_{\phi  }(\omega=0) = - ( p-1) J^p G_{F}(\omega=0)  \bar G_\phi^{p-2}   = - { 1 \over G_{\phi   }(\omega=0)  } = - { 1 \over { \beta  }  \bar  G_\phi }
 \ee 
 where we also imposed \nref{eq:SD_eqs_Sigma}. Using \nref{GFan} we get an equation purely for $\bar G_\phi$ which becomes 
%
 \be \la{ValSe}
 J {\bar G }_\phi = { 1 \over [(p-2) \beta J]^{1\over p-1} }  .
 \ee 
  We can check that the terms we have neglected are subleading.   
  
  This expression agrees with an expression we can get from the original theory applying classical thermodynamics. The potential for the bosons scales like $\phi^{ 2(p-1) } $, so we get 
 that  
 \be 
 Z \sim \int d \phi \exp\left( - \beta J^p \phi^{ 2 ( p-1) } \right)  \longrightarrow  J \langle \phi^2 \rangle \propto { 1 \over  (\beta J)^{  { 1 \over p -1 } } } 
   \ee 
 
   \subsection{Naive low energy analysis of the equations} 
   
   In this section we perform a naive analysis for the low temperature regime, leaving the correct analysis for  section \ref{LowEn}. From on now, we mostly set $J=1$.
   
   At low temperatures, we can attempt to solve the equations by making a scaling ansatz where the functions are taken to be 
   \be 
   G_{A} \propto {1 \over z^{2 \Delta_A} } ~,~~~~~~~A = \phi,~\psi,~F 
   \ee 
   and then inserting this into the equations, keeping the leading terms. 
   
   If we expect that supersymmetry is not broken then we would also require a certain relation between the different functions, which is 
   \be \la{SusRe}
   G_F = - \partial_\tau^2 G_\phi ~,~~~~~~~~G_\psi = - \partial_\tau G_\phi 
   \ee 
   This relation comes from applying the supercharges on the correlators and assuming that they annihilate the vacuum, at least to leading order in $N$. One might question this assumption based on our computation of the index for the ${\cal N}=2$ theory. However, the numerical solutions suggest that supersymmetry is not broken at this order in the large $N$ expansion. So this is a reasonable assumption. 
   
   This then implies that we can write an ansatz of the form 
   \be \la{Ansa}
   G_\phi = { c \over |\tau|^{2\Delta } } ~,~~~~~~~G_\psi = {\rm sgn}(\tau) {   2 \Delta c \over |\tau|^{2\Delta +1 } } ~,~~~~~~~G_F = - {   2\Delta(2\Delta +1 ) c\over |\tau|^{2\Delta + 2 } }
   \ee 
   We can insert this ansatz into the large $N$ equations  \eqref{eq:SD_eqs_Sigma} and \eqref{eq:SD_eqs_G_general_q}, neglecting the $1,\omega ,\omega ^2$ compared to $\Sigma_A $ in \eqref{eq:SD_eqs_Sigma}. We obtain an equation of the form 
    \begin{equation} \la{NlE}
 1 =  8\Delta (2\Delta +1) \sin (\pi (p-1)\Delta ) \sin ( \pi \Delta)  \Gamma \left( 1-2(p-1)\Delta \right) \Gamma (-1-2\Delta ) c^p |\omega |^{2p\Delta } .
 \end{equation}
 
   Since the left hand side is $\omega$ independent we conclude that $\Delta =0$ \cite{Anninos:2016szt}. 
   However, the prefactor then implies that the right hand side is zero.\footnote{Strictly speaking when $\Delta =0$ we do not get a power law in the equation but rather a $\delta (\omega )$, however the equation is still not satisfied.} Therefore this strategy does not lead to a solution. 
   
   However, let us imagine that $\Delta$ is very small and $c$ large in such a way that we expand \nref{NlE} in powers of $\Delta$ which gives 
 \begin{equation}
 1=4\pi ^2 c^p (p-1) \Delta ^2+\cdots \qquad \longrightarrow \qquad \Delta =\frac{1}{2\pi \sqrt{p-1}c^{p/2} } .
 \end{equation}
 
   We can then expand the above functions \nref{Ansa} keeping the first two terms in $G_\phi$. We write the analogous result at finite temperature  
    \bea \la{AnsaFT}
   G_\phi &=& \bar G_\phi - { 1 \over \pi \sqrt{p-1} \, {\bG}^{p-2\over 2}  } \log \sin{ \varphi \over 2}    ~,~~~~~~~~~~~~~~\varphi = {2\pi \tau \over \beta }
  \cr 
   G_\psi &=&   { 1 \over \sqrt{p-1} \, {\bG}^{p-2\over 2}   \beta \sin{ \varphi \over 2}   }~,~~~~~~~G_F = -\frac{\pi}{\sqrt{p-1}}{ 1 \over {\bG}^{p-2\over 2}   \beta^2 \sin^2 { \varphi \over 2}  }    \eea  
   where we have defined $c = \bG$ which would be the average value of $G_\phi$. We will later reproduce this from a more sophisticated analysis, which will also fix the value of $\bG$ which has not been fixed by the present analysis.

   \section{Numerical solution of the equations } \label{numsec}

Here we discuss a numerical solution of the Schwinger-Dyson equations \nref{eq:SD_eqs_Sigma}, \nref{eq:SD_eqs_G_general_q}. Similar equations were solved numerically in 
\cite{Kabat:1999hp}. There, the equations were obtained as a truncation of the full sum over planar diagrams. This truncation boils down to the restriction to melon diagrams.  Though these truncations are not justified for those models, the final equations are the same as those that describe the large $N$ limit of the SYK-like models discussed here. 
Similar BFSS inspired truncations were  solved numerically in \cite{Kabat:2000zv,Kabat:2001ve}.

We used the method of weighted iterations described in \cite{Maldacena:2016hyu}. Roughly, the idea is to begin with some initial functions for $G$ and $\Sigma$, then iteratively apply the equations of motion. This generates a sequence of improving approximate solutions. The procedure is continued until the desired convergence is achieved.
   
We will now describe the technical details of this method. The imaginary time interval $(0,\beta)$ is discretized into $n = 2^{15}$ points. Likewise, the bosonic and fermionic frequencies are taken in the ranges $\omega_{b} \in \lp -\frac{ \pi n}{\beta }, \frac{ \pi n}{ \beta }-\frac{2 \pi}{\beta}\rp$ and $\omega_{f} \in \lp -\frac{\pi n}{\beta} + \frac{\pi}{\beta}, \frac{\pi n}{\beta}-\frac{\pi}{\beta}\rp$, the interval between adjacent frequencies being $\frac{2 \pi}{\beta}$. We begin with some seed propagators $G^{(0)}_{A}(\omega)$, $A = \phi, \psi, F$. These could be, for example, the propagators of the free theory. The propagators are found in the time domain using a fast fourier transform. The self-energies $\Sigma_{A}^{(0)}(\tau)$ are then calculated using one half of the Schwinger-Dyson equations \nref{eq:SD_eqs_G_general_q} and fourier transformed to frequency space. These $\Sigma_{A}^{(0)}(\omega)$ determine propagators $G_{A}^{(1)}(\omega)$ via the other half of the equations \nref{eq:SD_eqs_Sigma}. Finally, the updated propagator is given by a weighted average of $G_{A}^{(0)}(\omega)$ and $G_{A}^{(1)}(\omega)$. For example, for the boson,
\begin{align}
	G_{\phi}^{(i)}(\omega) = (1-x)G_{\phi}^{(i-1)}(\omega) + \frac{x}{\omega^{2}-\Sigma_{\phi}(\omega)}, \quad \quad 0 < x < 1
\end{align}
and similarly for the other fields. $x$ is a weight that should be adjusted for convergence. This loop is executed until the approximate solutions stop changing up to some preset tolerance. The test for convergence is done by calculating
\begin{align}\label{conv}
	\int_{0}^{\beta} d \tau ~\big | G_{A}^{(j)}(\tau) - G_{A}^{(j-k)}(\tau)\big |^{2}
\end{align}
between successive iterations for some integer $k$. The test for convergence can be made stricter by choosing larger $k$ or a smaller upper bound on \nref{conv}. Another quantity that can be useful as a measure of convergence is the change of the boson zero mode between iterations,
\begin{align}
	\frac{1}{\beta} \lp \bigg | \int_{0}^{\beta} G_{\phi}^{(j)}(\tau)d\tau \bigg | - \bigg | \int_{0}^{\beta} G_{\phi}^{(j-k)}(\tau)d\tau \bigg | \rp
\end{align}
This was the method used to calculate the numerical solutions shown in Figures \ref{pl1}, \ref{lgZplt}, and \ref{fnsplt}.

We also plot analytic solutions against the numerical solutions for comparison. The high-temperature analytic solutions were derived in section \ref{HighEn}, and the low-temperature analytic solutions will be derived later in section \ref{LowEn}.

\begin{center}
\includegraphics[height=8cm]{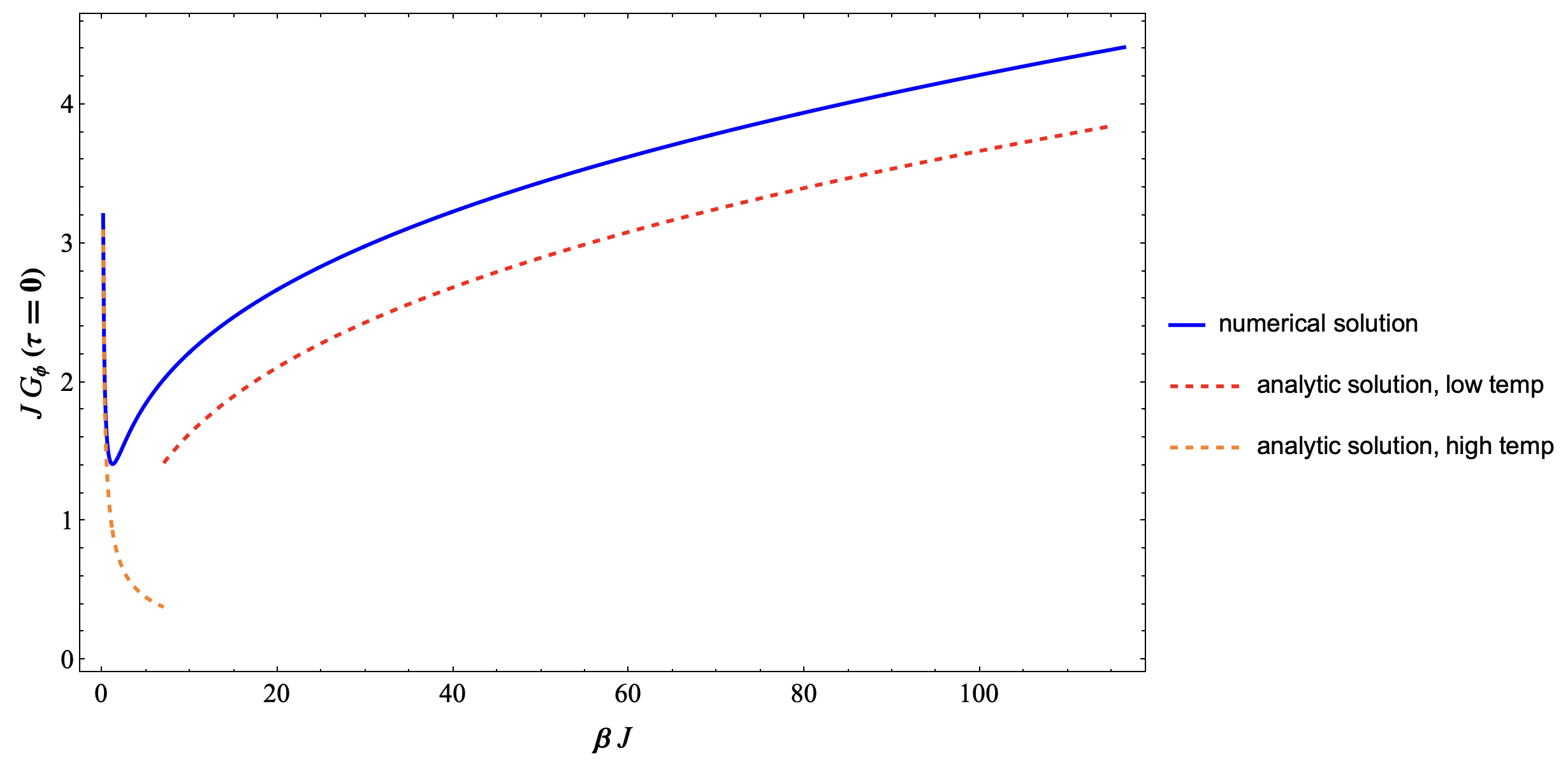}
\captionof{figure}[Numerical solution for $J G_{\phi}(\tau =0)$ in the $p=3$ model, plotted against the high-temperature \nref{ValSe} and low-temperature \nref{PhiSq} analytic predictions.]{Numerical solution for $J G_{\phi}(\tau =0)$ in the $p=3$ model, plotted against the high-temperature \nref{ValSe}\footnotemark\  and low-temperature \nref{PhiSq} analytic predictions.
The difference  between the analytic and numeric answers for large $\beta J$ could be explained by including higher order terms in the effective action.  What is important is that the fractional difference is decreasing. }
\label{pl1}
\end{center}
\footnotetext{\nref{ValSe} gives a high-temperature expression for $J G_{\phi}(\omega = 0)/\beta$, while $J G_{\phi}(\tau = 0)$ includes contributions from nonzero frequency modes as well. However, at high temperatures the nonzero modes can be neglected, so $J \bar G_{\phi}$ is a good approximation to $J G_{\phi}(\tau = 0)$.}

\begin{center}
\includegraphics[height=8cm]{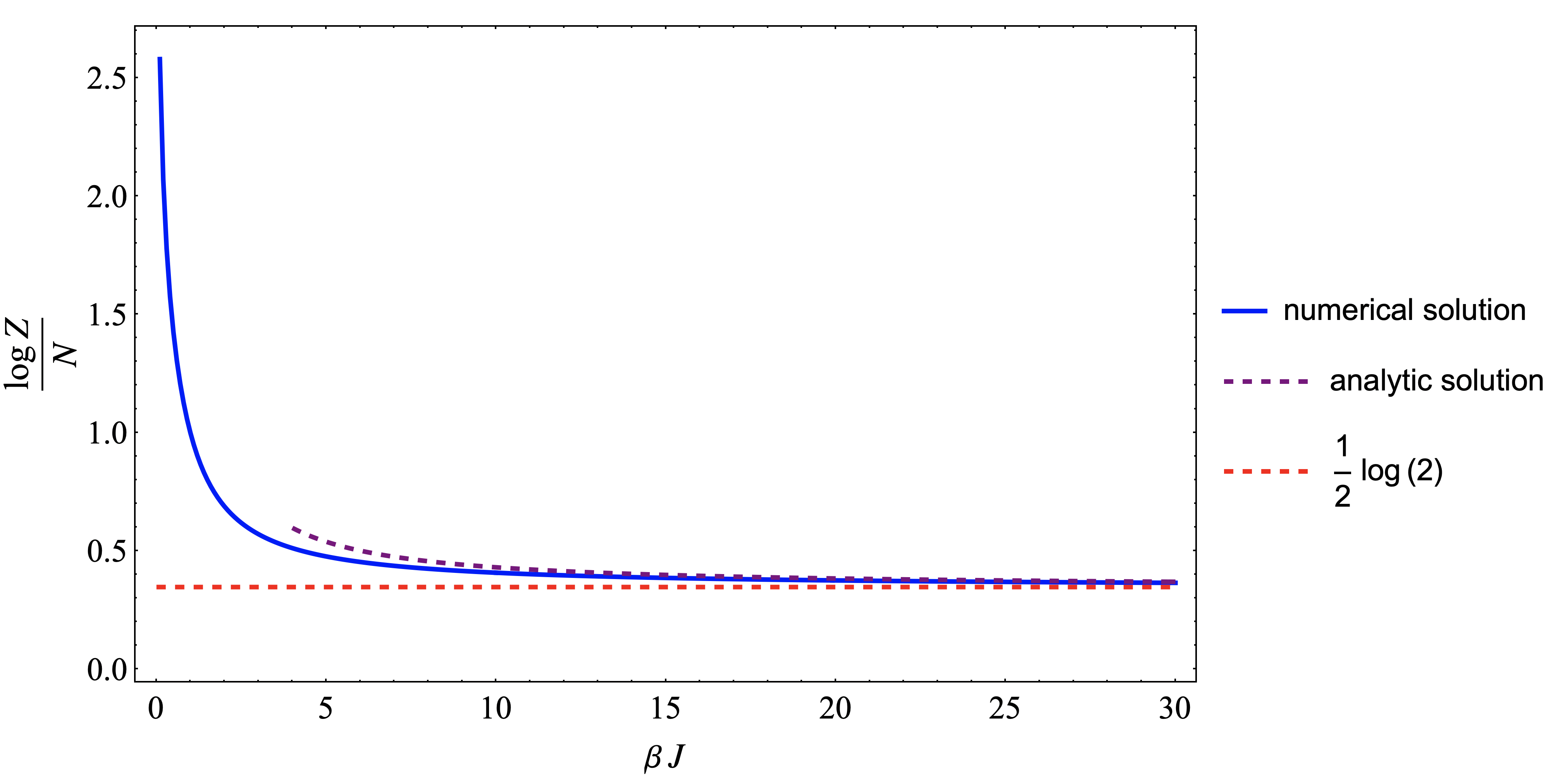}
\captionof{figure}{Numerical solution for $\frac{\log Z}{N}$ in the $p=3$ model, plotted against the low-temperature analytic prediction \nref{FinAcG}. We see that it asymptotes to the ground state entropy term $S_{0} = \frac{1}{2}\log 2$ of the large $N$ effective action. The energy, given by minus the slope of $\log Z$, goes to zero in the low temperature limit at this order in the $1/N$ expansion. }
\label{lgZplt}
\end{center}
\centerline{
\includegraphics[width=.64\textwidth]{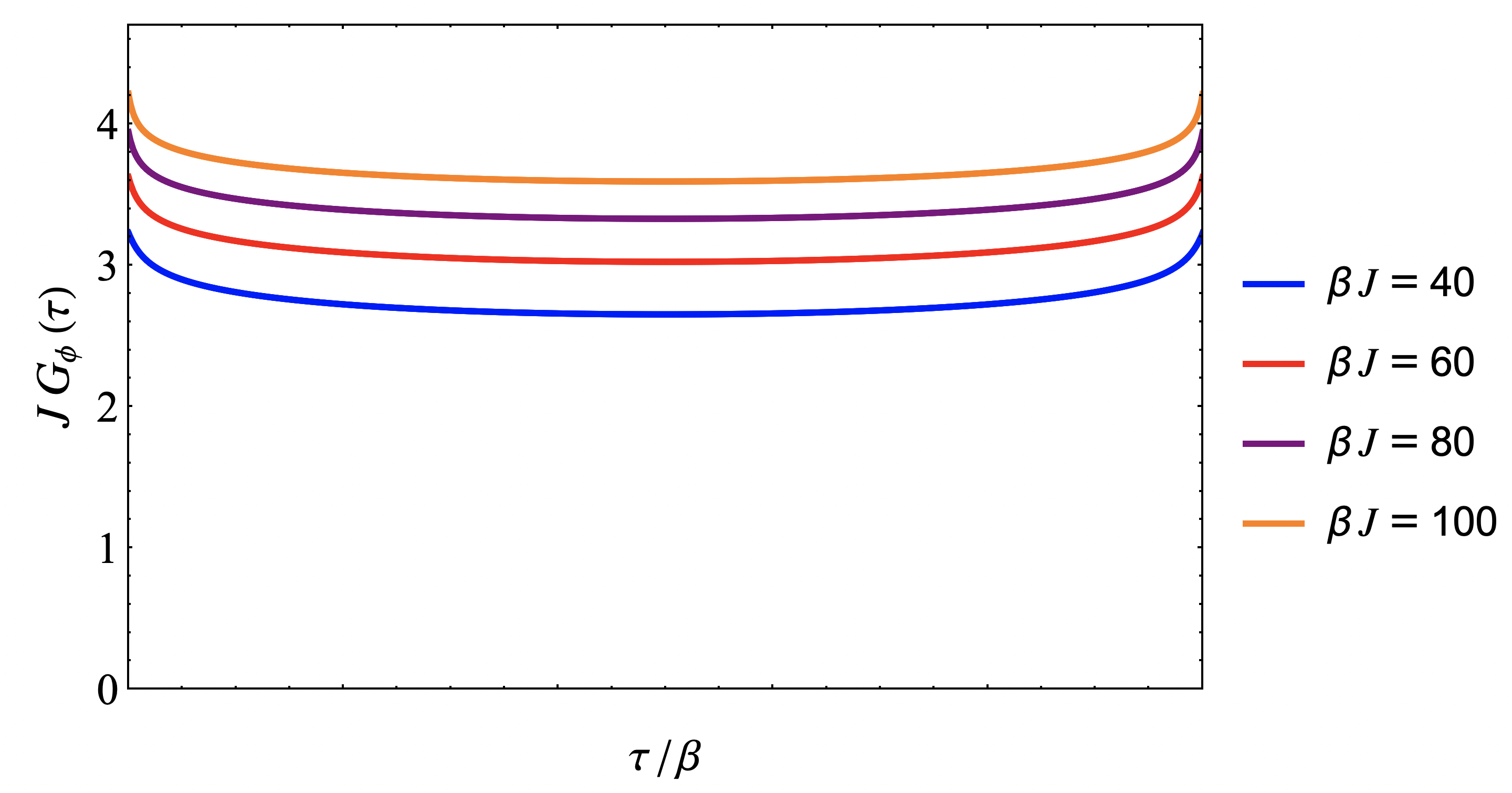}\hfill
\includegraphics[width=.64\textwidth]{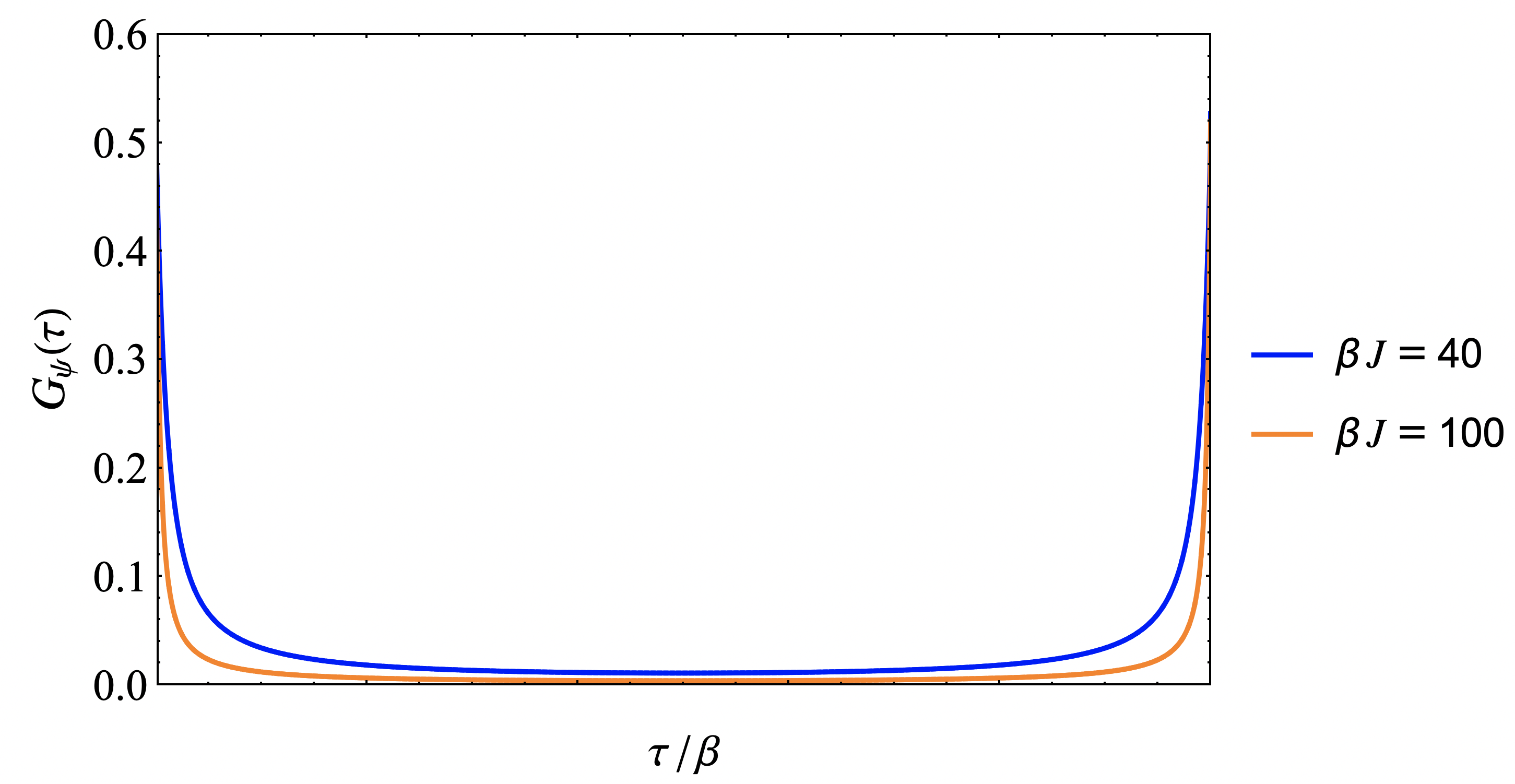}\hfill}

\begin{center}
\includegraphics[width=.66\textwidth]{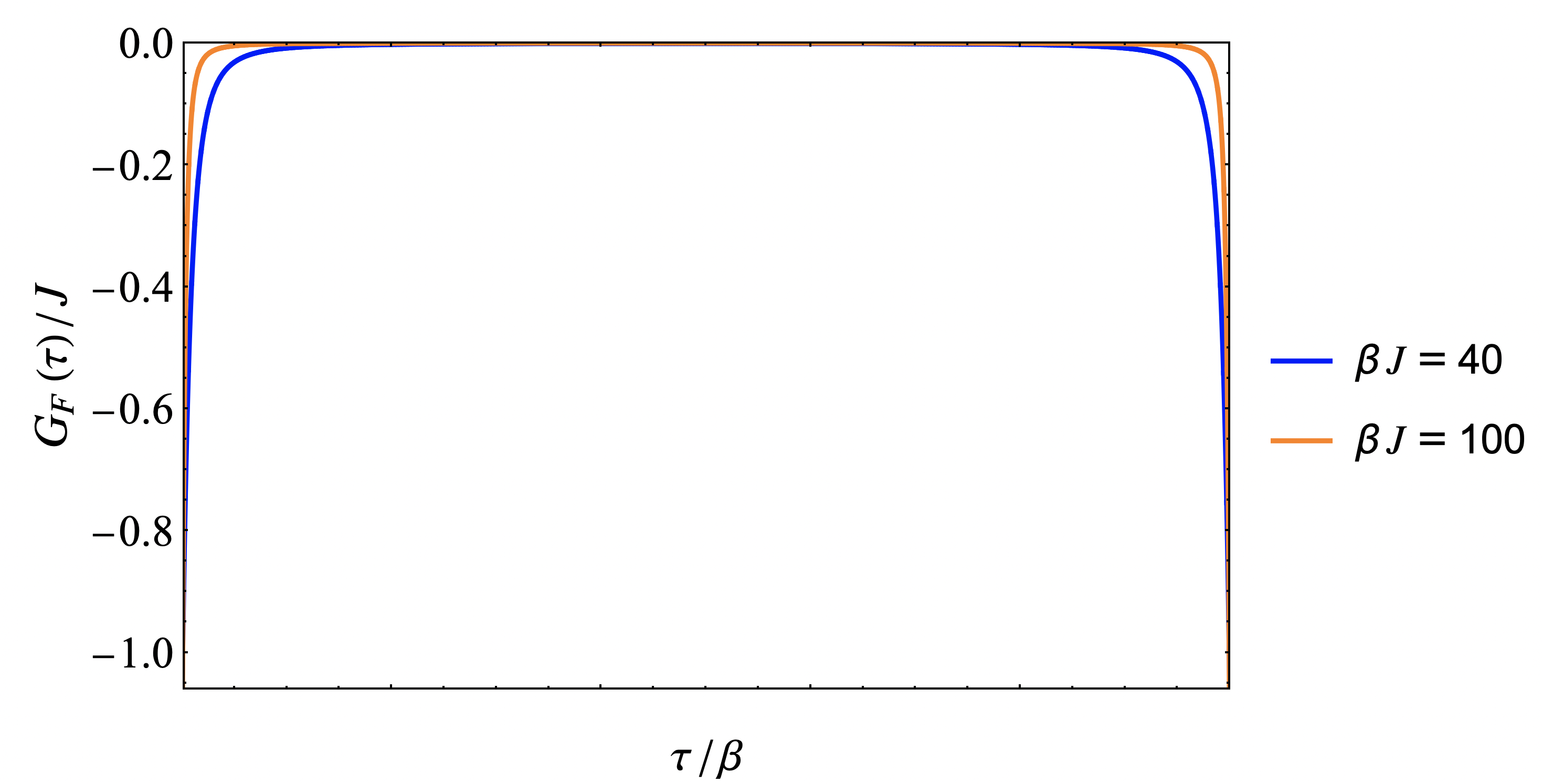}
\captionof{figure}{Numerical solutions for the Euclidean two-point functions $JG_{\phi}$, $G_{\psi}$, and $G_{F}/J$ at various $\beta J$ in the $p = 3$ model.  }
\label{fnsplt}
\end{center}

   \section{Low temperature analysis of the equations } 
   \la{LowEn}

   The analytic approximation to the small temperature analysis is somewhat involved and so we devote this whole section to it. This analysis was originally done in \cite{Lin:2013jra} for $p=3$. Here we rederive that in a slightly more direct way and extend it to any $p$.  We set $J=1$, and we can restore it by dimensional analysis. 
   
   We will do a series of approximations at the level of the effective action. The main idea is to say that the function $G_\phi$ becomes a large constant value denoted by $\bar G_\phi$ plus something small that represents its non-zero Matsubara frequency contributions.  
    We will expand in these corrections perturbatively. This is motivated by looking at the numerical solutions in figure \ref{fnsplt}.  Since the discussion is somewhat long, let us just state that   we will find that the effective action has the form 
   \be 
   \log Z = S_0 + { c_1 \over \beta \bG^{ p-2 \over 2} } - { c_2 \over \bG^{p } } ~,~~~~~\to ~~~\bG \propto \beta^{ 2 \over p +2} \la{SolSche}
   \ee 
   where $S_0$, $c_1$ and $c_2$ are some numbers. We have also indicated the scaling with temperature of the value of $\bG$ that maximizes this function. 
   
   \subsection{Rewriting the effective action} 
   
The original effective action \nref{EAFirst} can be written as 
\bea \la{OrAct}
{\log Z \over N } &=& \sum_{\omega_n} \left[ \log[ - i \omega  -   \Sigma_\psi  ] - 
\half \log (\omega^2 -  \Sigma_\phi) - 
\half \log ( 1 -  \Sigma_F) \right] + \cr  
& ~& + \int_0^\beta  dt dt' \left( -\half  G_\phi \Sigma_\phi - \half G_F \Sigma_F -   G_\psi \Sigma_\psi - \half G_F G_\phi^{p-1} + { (p-1) \over 2} G_\psi^2 G_\phi^{p-2} \right)~~~~~~~~~~~~~
\cr 
&~& ~
\eea
where the first line is in frequency space and the second in time. The frequencies of the bosons and fermions are 
\be \la{FreqBF}
\omega_n = { 2\pi n \over \beta } ~,~~~~~\omega_r = { 2 \pi (n+\half) \over \beta }
\ee 
We can also rewrite it as 
\bea \la{OrAct2}
{\log Z \over N }  &=& 
\sum_{\omega } \left[ \log[ - i \omega  -   \Sigma_\psi (\omega)] - 
\half \log (\omega^2 -  \Sigma_\phi) - 
\half \log ( 1 -  \Sigma_F) \right.
 + \cr  
 & ~&   \left. 
 - \half G_\phi \Sigma_\phi - \half G_F \Sigma_F +   G_\psi(\omega) \Sigma_\psi(\omega) \right] +
\cr 
&~&  +\int d\tau d\tau' \left( - \half G_F G_\phi^{p-1} +{ (p-1) \over 2} G_\psi^2 G_\phi^{p-2} \right)~\eea
where now the first two lines are in frequency space.\footnote{Note that the bosons and the fermions are summed over different frequencies, see \nref{FreqBF}.} 
Note that for a quadratic expression the double integral over time is equivalent to the sum over frequencies. 
We have also used that for the fermions $G_\psi(-\omega) =- G_\psi(\omega)$ to rewrite it in terms of functions all evaluated at $\omega$. The sums are over both positive and negative (and zero) frequencies.  

It is convenient to integrate out the $\Sigma_A$ variables, here $A=\phi,F,\psi$. 
The equations for $\Sigma_A$ come  from the first two lines of \nref{OrAct2}  
\be 
 0= { 1  \over - i \omega -   \Sigma_\psi} -  G_\psi =  { 1 \over \omega^2 -  \Sigma_\phi } - G_\phi=  { 1 \over 1 -   \Sigma_F} - G_F =0
\ee 
These equations allow us to solve for $\Sigma_A$ in terms of $G_A$. We now replace those solutions back into the   action \nref{OrAct2} to obtain 
 \bea \la{AcTG}
{\log Z \over N } &=& \sum_{\omega }\left[  - \log G_\psi + \half \log G_\phi G_F  - { \omega^2 \over 2 } G_\phi - { \half } G_F  - i \omega G_\psi  \right]+ \cr  
& ~& + \int_0^\beta  dt dt' \left(  - \half G_F G_\phi^{p-1} + { (p-1) \over 2} G_\psi^2 G_\phi^{p-2} \right)~~~~~~~~~~~~~
\eea
 When we make this substitution, we encounter a constant of the form 
\be \la{SumOnes}
 \sum_{\omega_n} 1  - \sum_{\omega_r} 1 = \lim_{\epsilon \to 0 } \left[  \sum_{n} e^{ -|n| \epsilon }     - \sum_{n}   e^{ - \epsilon |n+\half | }  \right] =   0 
 \ee 
 where we have that the constant is zero in the limit that the regulator $\epsilon \to 0$.

 The new action \nref{AcTG} is equivalent to the original one, classically.

 \subsection{Expansion to quadratic order } 
 
 We write 
 \be \la{Gexpn}
 G_\phi = \bG + \delta G_\phi 
 \ee 
 where $\bG$ is the zero frequency part and $\delta G_\phi$ is the non-zero frequency part. We will assume that $\bG \gg \delta G_\phi, ~G_\psi ~,~G_F $ and expand the action in $\delta G_\phi$. 
Note that the first line in \nref{AcTG} is already   in frequency space. The expansion will 
  consist in expanding the second line in \nref{AcTG}   in powers of the non-zero frequency modes. The first non-trivial term is quadratic in the non-zero frequency modes.  
More explicitly, we write  the second line of \nref{AcTG} as 
 \bea 
   & ~&  -\beta \half G_F(\omega =0) \bG^{p -1} + V_2 ~,~~~~~~{\rm with:}
    \cr 
  V_2 &=&    \half \alpha \beta \int d\tau \left[ -\delta G_F(\tau) \delta G_\phi(\tau)  + G_\psi(\tau)^2 \right] = \half \alpha \sum_{\omega} \left[ -\delta G_F(\omega) \delta G_\phi(\omega)  - G_\psi(\omega)^2 \right] ~~
  \cr 
  ~~~ &~ &  \alpha \equiv   (p-1) \bar G^{p-2}_\phi \la{QuadTe} 
   \eea 
 where in the second line we used a frequency space expression and  that $G_\psi(-\omega) = -G_\psi(\omega)$.  
  
 We now solve the equations for this truncated theory. Just to be completely explicit, the truncated theory is 
 \bea \la{AcTGtr}
{\log Z_2 \over N } &=& \sum_{\omega } - \log G_\psi + \half \log G_\phi G_F  - { \omega^2 \over 2 } G_\phi - { \half } G_F  - i \omega G_\psi   \cr  
& ~& -\beta \half G_F(\omega =0) \bG^{ p-1}  + \half \alpha \sum_{\omega\not =0} \left[ -\delta G_F(\omega) \delta G_\phi(\omega)  - G_\psi(\omega)^2 \right] ~~~~~~~~~~~~~~
\eea
 Here, $\delta G_F , ~\delta G_\phi, ~G_\psi$ are the non-zero frequency components of the corresponding fields. Of course, $G_\psi$ has only non-zero frequencies. 
We consider the equations for all fields except for $\bar G_\phi$. 
The zero mode equation gives 
\be \la{ValGf}
0={ 1 \over G_F(0) } -1 -\beta \bG^{p-1} \to ~~G_F(0) = { 1\over 1 + \beta \bar G_\phi^{p-1} } \sim { 1 \over \beta \bG^{p-1} } + \cdots  ~,~~~~~\bar G_F = { G_F(0)\over \beta } \sim { 1 \over \beta^2 \bG^{p-1} }
\ee
The non-zero frequency mode equations are 
\be \la{EomQu}
- { 1 \over Q_\psi } - i \omega - \alpha Q_\psi =0 ~,~~~~{ 1 \over Q_\phi } - \omega^2 - \alpha Q_F=0 ~,~~~~~ { 1 \over Q_F } - 1 - \alpha Q_\phi =0
 \ee
 where $Q_A$ denotes the classical solutions of the truncated action for  $\delta G_\phi ~,\delta G_F, ~G_\psi$ for $A=\phi, F, \psi$. The solutions are
 \bea \la{SolG}
  Q_\psi &= & { 1 \over \sqrt{\alpha } }   i H(\omega ) ~,~~~~~H(\omega) \equiv  \left[- {| \omega | \over 2 \sqrt{\alpha} } + \sqrt{ 1+ {\omega^2 \over 4 \alpha } } \right] 
\cr 
   Q_F &=&   { 1 \over \sqrt{\alpha } } |\omega | H(\omega)    \cr
 Q_\phi &=&    { 1 \over \sqrt{\alpha } } { 1 \over | \omega |} H(\omega)    ~~~~~~~~~~~~~~\alpha \equiv   (p-1) \bar G^{p-2}_\phi
\eea

These expressions are for $\omega>0$, for $\omega< 0$ the fermion one gets a minus sign. For large $\omega$, $H \sim  { \sqrt{\alpha} \over | \omega |} $, which reproduces the expected short distance behavior. 
  
We now evaluate the action  to obtain 
\bea \la{Acqu}
 { \log Z_2 \over N } &=& - \half \log \bar G^{p-2}_\phi  + \half +  \sum_{w_r > 0 } - 2 \log Q_\psi - {  \alpha }  (i Q_\psi)^2    
 \cr 
& ~& + \sum_{w_n>0}     \log Q_F Q_\phi  + 
      {   \alpha }  Q_F Q_\phi 
 \eea
Let us explain the various contributions to this formula.   The term involving $\log \bG^{p-2}$ comes from the zero frequency part of $\log G_F G_\phi $ and \nref{ValGf}. We   neglected terms going like $1/( \beta \bar G_\phi^{p-1})$ which will be subleading. 
  We also restricted the sum to be over positive frequencies.  In deriving \nref{Acqu}, we used the equations of motion \nref{EomQu}. In particular, the linear terms in \nref{AcTG} give a contribution that is constant and also a contribution that is quadratic in $H$ that combines with the quadratic term in $H$ that comes from the second line of \nref{AcTG}. 
  The sum over all the  constant contributions can be done as follows, we get a $(-1) $ from the non-zero frequency bosons and a 1 from the fermions. The zero frequency term just gives a $-\half $ from   inserting \nref{ValGf} into the first term of the second line of \nref{AcTGtr}. These terms then give 
   \be 
  -\half +  \sum_{\omega_n \not =0} (-1) + \sum_{\omega_r } 1 = \half 
  \ee 
  where we used \nref{SumOnes}. This is the origin of the half in \nref{Acqu}.  
  
We can rewrite \nref{Acqu} as 
 \bea 
 { \log Z_2 \over N } &=& - \half \log \bar G^{p-2}_\phi  + \half  - \half f(0) + \sum_{n=0}^\infty [ \half f(n+1) + \half f(n) - f(n+1/2) ] 
 \cr 
f(n) & =& - \log \alpha +  2 \log h(n) + h(n)^2 ~,~~~~~
\cr 
 h(n) &\equiv& \hat H\left( { 2 \pi n \over \beta \sqrt{\alpha }} \right) ~,~~~~~~~\hat H(x) = - \half x + \sqrt{1 + {x^2 \over 4} }  \la{Z2Log}
 \eea 
 We can now evaluate  each sum by   using the Euler-Maclaurin formula 
\bea 
\sum_{n=0}^\infty  f(n) &=& \int_{n=0}^\infty dn f(n) + \half f(0) - { 1\over 12 } f'(0) + \cdots 
\eea
In our case, this becomes 
\bea \la{Sumsge}
\sum_{n=0}^\infty [ \half f(n+1) + \half f(n)  - f(n+\half) ] &=& \half \int_0^{\half} dn f(n) -\half \int_{1/2}^1 f(n) dn + \half \left[\half f(1) + \half f(0) - f(\half )\right] + \cdots 
\cr
&=&   - { 1\over 8 } f'(0) 
\eea
%
Using now that 
 \be f(0) = - \log \alpha +1 ~,~~~~~~f'(0) = -2 \times { 2 \pi \over \beta \sqrt{\alpha } } 
 \ee 
we find that 
   \be \la{ZPartAn}
  { \log Z_{2} \over N } =  \half \log(p-1)     + {   \pi \over 2} { 1 \over \beta \sqrt{ (p-1) }\bar G_{\phi}^{{p\over 2} -1}  } + \cdots 
  \ee 
  Note that the terms involving $\log \bar G^{p-2}_\phi$ cancel out. We see that the first term looks a bit like a ``zero temperature'' entropy.  Note that, for fixed $\beta$, the second term  wants to make $\bG$ smaller. In section \ref{PhInterp},  we discuss a physical interpretation of these two terms. 

\subsubsection{Side comment on $p=2$}\la{Ptwo}
 
 This derivation is correct for $p>2$. For $p=2$ we should analyze it separately. In that case we do not want to make the last approximation in \nref{ValGf} and the first log in \nref{Acqu} gets replaced by 
 \be \la{Qtwo}
{\log Z \over N } \bigg|_{p=2} =  \half \log \left( { \beta \bar G_\phi \over 1 + \beta \bar G_\phi } \right) +  { \pi \over 2 \beta } 
 \ee  
   Note that for $p=2$, the second term, which comes from  the non-zero frequency part, is completely independent of $\bG$. 
   If we want to maximize  \nref{Qtwo} with respect to  $\bG$ we need to send it to infinity. As  $\bar G_\phi \to \infty $,   the logarithm becomes zero and the action remains finite.   We will give a physical explanation of this behavior in section \ref{PhInterp}.

 \subsection{Higher order terms} 

 There are several higher order terms. Many of them give the same type of answer. In fact, they will boil down to essentially the same expression. This subsection is somewhat technical, so the impatient reader might want to jump directly to section
 \ref{Collecting}.

 \subsubsection{Subleading quadratic terms} 
 
 The first term has the form 
  \bea 
  V_{2'} &\equiv &   d_3  \beta \int d\tau \bar G_F\delta G_\phi^2 = d_3 \bar G_F\sum_{\omega }  Q_\phi(\omega )^2 
  \cr  
  &~&~~~{\rm with} ~~~d_3  \equiv   -{ (p-1)(p-2) \over 4 }  \bG^{p-3} ~,~~~~~~~\bar G_F = { 1 \over \beta^2 \bG^{p-1} }  \la{D3def}
 \\ 
 V_{2'} &=& d_3 \bar G_F { 1 \over \alpha } {\beta^2 \over (2 \pi)^2} 2 \sum_{n} { 1 \over n^2 } = { d_3 \bar G_F \beta^2 \over \alpha } { 1 \over 12 } 
 \cr 
 V_{2'} &=& - { (p-2) \over \bar G_\phi^p } { 1 \over 48 }  \la{V2pr}  \eea
where we approximated $H\sim 1$ when we substituted the solution \nref{SolG}.

\subsubsection{The cubic terms } 

We now turn our attention to a term of the form 
\bea 
V_3  &=& d_3 \beta \int d\tau (\delta G_F \delta G_\phi - 2 G_\psi^2 ) \delta G_\phi \la{CubTe}
\eea
with $d_3$ as above,  \nref{D3def}.  

The first contribution would come from  evaluating this term on the previous solution \nref{SolG}. 
For that purpose, it is important to note some aspects of the quadratic solution \nref{SolG}. 

 We define a new fermion function $\tilde Q_\psi$ as similar to $Q_\psi$ \nref{SolG} but with integer frequencies. Then we have the following relations between the functions 
 \be \la{susyRel}
 Q_F = - \partial^2 Q_\phi ~,~~~~~~ \tilde Q_\psi = - \partial Q_\phi 
 \ee 
 This is the relation we would have between the different functions if we had a supersymmetric solution. Because of the finite temperature,  the fermion solution is $Q_\psi$ which has half integer frequencies. 
It is important to understand more clearly the relation between $Q_\psi$ and $\tilde Q_\psi$.  We will need this relation in Euclidean time space and only for large $\alpha$, where we approximate $H\sim 1$. 
 In that limit we have 
 \bea \la{LsQ}
 Q_\phi &=& -{ 1 \over   \sqrt{\alpha} } { 1 \over 2 \pi } \log[ 4 \sin^2 {\varphi\over 2 }  ] ~,~~~~~~\varphi \equiv { 2 \pi \tau \over \beta } 
 \cr 
 \tilde Q_\psi &=&  - \partial_\tau Q_\phi = { 1 \over   \sqrt{\alpha}\beta   } { \cos{\varphi \over 2} \over \sin{\varphi \over 2 }}  ~,~~~~~~~~~~~~
 Q_\psi = { 1 \over \sqrt{\alpha}\beta  } { 1 \over \sin{\varphi \over 2 }}  
 \cr 
 Q_F & =& - { 1 \over  \sqrt{\alpha} } { 2 \pi \over \beta^2} \half { 1 \over [ \sin {\varphi \over 2}]^2 } 
  \eea
  where we assume that $\varphi$ is not too small. We will   use the following property 
  \be \la{CosSin}
  Q_\psi^2 - \tilde Q_\psi^2 = { 1 \over \beta^2 \alpha } 
  \ee 
  This is an equation valid to leading order in the 
  $1/(\beta \sqrt{\alpha} ) $ expansion. 

By the way, note that \nref{Gexpn} together with the solutions \nref{LsQ} agree with what we had from the naive analysis in \nref{AnsaFT}. 

 \subsubsection{The cubic term on the solution} 
 
 We can now evaluate the cubic term \nref{CubTe} on the solution \nref{SolG}. 
 Let us first estimate the order of this term. We expect 
 \be 
     \bG^{p-3} { 1 \over \alpha^{3/2}} { 1 \over \sqrt{\alpha } } \sum { 1\over n }  \propto { 1 \over \bG^{ {p \over 2}  }} \left( 1 + { 1\over \beta \sqrt{\alpha } } \right) 
 \ee 
 where the last sum is schematic. We see that the leading term is more dominant than the terms we have kept, after we take into account the final scaling with $\beta $ of the solution \nref{SolSche}. However, we will show that this leading term is actually zero. Furthermore, the subleading terms are less important than the other terms we kept. So we just need to show that the leading order term is zero. 
 
 In order to show this, we can simply use the leading solutions with $H\sim 1$ given in \nref{LsQ}. 
 We will rewrite the cubic term as 
 \bea 
 V_3 &\propto  &   \int d\tau (Q_F Q_\phi - 2 Q_\psi^2)  Q_\phi = \int d\tau (Q_F Q_\phi - 2 \tilde Q_\psi^2)  Q_\phi +  2 \int d\tau   (\tilde Q_\psi^2 -   Q_\psi^2)  Q_\phi = 
 \cr 
 & \propto  & - \int d\tau [ \partial_\tau^2 Q_\phi Q_\phi + 2 (\partial_\tau Q_\phi)^2]  Q_\phi  - { 2 \over \alpha \beta^2} \int Q_\phi 
 \cr 
 & \propto & - \int d\tau \partial_\tau ( Q_\phi^2 \partial_\tau Q_\phi)  =0  \la{CubTD}
  \eea
 where we used \nref{CosSin} for the second term, and we used that the integral $\int Q_\phi =0$, since the $Q$s contain only non-zero frequencies. We also rewrote the first term as a total derivative, which vanishes since it is the derivative of a periodic (but discontinuous) function   on the circle\footnote{The discontinuity arises from the short distance behavior in 
 \nref{ShortD}. This leads to a $\delta$ function that should be included in \nref{CubTD} in order to apply this total derivative argument.}. 
 
 \subsubsection{The quadratic fluctuations and the cubic term} 
 
 There is a second contribution from the cubic term that is easy to miss. It   arises because the cubic term is formally somewhat large, so that further corrections are important. Let us discuss these corrections in detail. 
 
 Just to get oriented  with the type of problem we have, let us just imagine we do an integral over a single variable $G$ 
 \bea 
  &~& \int d G \exp\left[ - \half (G- Q)^2 + {1\over 3 } d_3 G^3 \right] 
  \cr 
  &=& \int d g \exp \left[- \half g^2 + {1\over 3} d_3 Q^3 +   d_3 Q^2 g \right] ~,~~~~~~~G= Q + g
  \cr 
  & = & \exp\left[{1 \over 3}  d_3 Q^3 + ( d_3 Q^2 )^2 \right] 
 \eea 
 where we say that the quadratic solution is just $G=Q$. Then we expand around it for the cubic term. We have just indicated the first leading term, the $Q^3$ term,  and the first subleading term, which is the only one we will need in our case. 
   
   We will now perform this expansion in detail for our theory. 
 Let us expand all the non-zero frequency modes as 
   \be \la{Insq}
   \delta G_A = Q_{A }  + g_A ~,~~~~~~~A = \phi, ~F ,~ \psi 
   \ee 
   where $g_A$ will be the fluctuations around the solutions.  
   
   Expanding the non-zero frequency terms in the action \nref{AcTG} we get  
     \be 
  {\log Z \over N } |_{\rm quad } \sim \sum_{\omega \not =0} ~
   { 1 \over2  Q_\psi^2} g_\psi^2 - { 1 \over 4   Q_\phi^2 } g_\phi^2  - { 1 \over 4   Q_F^2 } g_F^2 - { \alpha \over 2} ( g_\phi g_F + g_\psi^2 ) \la{QuaEx}
  \ee 
 Note that these are frequency space expressions. 
 Let us just insert the leading order expression for $Q_A$ (setting $H=1$ in \nref{SolG}).  
       \be \la{QuaGau}
  {\log Z \over N } |_{\rm quad } \sim \sum_{\omega \not =0} ~  - { \alpha  \over 4 } \left( |\omega|  g_\phi  + { 1 \over |\omega|  } g_F  \right)^2
- \alpha g_\psi^2   
  \ee 
 We   find that the bosonic quadratic terms have a zero mode. This disappears once we consider the next order correction. So, we need to consider the next order terms expansion for the $Q_A$ in 
 \nref{QuaEx}. 
  
 We can diagonalize the kinetic term by choosing the fields as 
 \be \la{DiagVa}
 g_\phi = { 1 \over |\omega| } \rho + { 1 \over |\omega| } \gamma ~,~~~~~~g_F = {  |\omega| } \rho - {  |\omega| } \gamma 
 \ee 
 With these definitions, the quadratic terms in the action become
 \be \la{QuaGau2}
  {\log Z \over N } |_{\rm quad } \sim \sum_{\omega \not =0} ~  - \alpha \rho(\omega)^2 - \half \sqrt{\alpha } |\omega | \gamma(\omega )^2  - \alpha g_\psi^2   
  \ee 
 (We neglected an extra term going like $\rho^2 |w| \sqrt{a}$ which leads to  subleading corrections). We see that the $\gamma^2$ term has a coefficient which is smaller than that of the $\rho^2$ or $g^2_\psi$ terms by a factor of ${ |\omega| \over \sqrt{\alpha }}$. 
 
 We now expand the cubic term to leading order in the $g_A$ fields. 
  We first start with the fermionic terms involving $g_\psi$. 
 Expanding \nref{CubTe} to leading order in $g_\psi $ we find 
 \be {\log Z_3 \over N } |_{\rm fer-linear} = - 4 d_3 \beta   \int dt Q_\psi Q_\phi g_\psi 
 \ee 
 This implies that the equation of motion for $g_\psi$ in time (as opposed to frequency space, beware of a change in sign) 
 \be 
 { 2 \alpha  } g_\psi(\tau) - 4 d_3 Q_\psi(\tau) Q_\phi(\tau) =0
 \ee 
 Inserting this back into the action we find 
 \bea \la{SigCo}
 {\log Z_{3 ~{\rm fer}} \over N } & =&  -  {4 d_3^2     \over \alpha } \beta \int d\tau Q_\phi^2 Q_\psi^2 
 \eea 
 This last integral appears to be divergent at short distances. It will combine with the bosonic part to give something finite. 
 
 Let us now discuss the bosonic part. Expanding the cubic term, \nref{CubTe},  using  \nref{DiagVa}, to leading order we find 
   \be 
 { \log Z_{3~{\rm bos-linear}} \over N } = d_3 \sum_{\omega\not =0}    \left\{ 
 \left[ 2 Q_F Q_\phi - 2 Q_\psi^2  + \omega^2 Q^2_\phi\right] (\omega) { 1 \over |\omega |} \rho(\omega) + 
 \left[ 2 Q_F Q_\phi - 2 Q_\psi ^2 - \omega^2 Q^2_\phi\right] (\omega) { 1 \over | \omega | } \gamma(\omega) \right\}
\ee
Where we take the product of the $Q$s in position space and then Fourier transform them to frequency space. It is important that the $\omega^2$ in this expression is the $\omega^2 $ of the perturbation, $\rho$ or $\gamma$, or alternatively the $\omega^2$ of the product $Q_\phi^2$. 

As we have done before,   we add and subtract terms with $\tilde Q_\psi^2$, which satisfy the  relations \nref{susyRel}. 
Let us consider first the terms that couples to $\gamma$. 
After that substitution we get 
\bea 
&\propto &   \left[ 2 Q_F Q_\phi - 2 \tilde Q_\psi^2 +  \partial_\tau^2(  Q^2_\phi) \right] (\omega) { 1 \over | \omega |  } \gamma(\omega)  + \left[    2 (\tilde Q_\psi^2 - Q_\psi^2 )  \right] (\omega) { 1 \over | \omega | } \gamma(\omega) 
\cr 
&\propto & \left[ - 2 \partial^2_\tau Q_\phi Q_\phi - 2 ( \partial_\tau Q_\phi)^2 + \partial_\tau^2 (Q_\phi)^2 \right] (\omega) { 1 \over | \omega |  } \gamma(\omega) -{ 2 \over \beta^2 \alpha} \left[    1  + o ( {|\omega | \over \sqrt{\alpha } } )  \right] (\omega) { 1 \over | \omega |  } \gamma(\omega) 
\cr 
& \propto & 0 
\eea
The term involving the derivatives cancels exactly. For the other term, the one involving 
 $(\tilde Q_\psi^2 - Q_\psi^2 )$, we used that, to leading order, it becomes just a constant, therefore it only has a zero frequency component. But we are considering non-zero frequencies here. We will discuss the zero frequency modes later. 
In principle, there are non-zero terms at subleading order. Given the fact that the original action for $\gamma$ has an enhancement, we need to worry about subleading terms. However, the enhancement is just a factor of ${ \sqrt{\alpha } \over \omega} $. But, since the leading order linear coupling to $\gamma$ cancels, it means that we would get {\it two} factors of ${ |\omega | } \over \sqrt{\alpha } $ which would overwhelm the enhancement, so that we do not get any contribution to the order we are working.  
The conclusion is that the terms involving $\gamma$ give no contribution. 

Now we consider the terms involving $\rho$. 
We perform a similar analysis, but now it is enough to consider everything to leading order. For the same reason as before, we can substitute $Q_\psi^2$ by $\tilde Q_\psi^2$. The error we make could only contribute to zero frequencies as before. 

The expression can be written as a total derivative, so that we get 
\bea 
 { \log Z_3 \over N }|_{\rm bos-linear} &= &\sum_{\omega\not =0}  d_3 \left[ 2 Q_F Q_\phi - 2 \tilde Q_\psi^2  + \omega^2 Q^2_\phi\right] (\omega) { 1 \over | \omega |  } \rho(\omega)  
 \cr 
 &= &\sum_{\omega\not =0}  -d_3
 \left[-  2 \partial^2   Q_\phi  Q_\phi - 2   ( \partial  Q_\phi)^2  - \partial^2  Q^2_\phi\right] (\omega) { 1 \over | \omega | } \rho(\omega)
 \cr 
 &= & \pm i d_3 \sum_{\omega\not =0}      
 \left[-  4 \partial   Q_\phi Q_\phi  \right] (\omega)  { \omega \over |\omega| }  \rho(\omega)
\eea
the factor of ${ \omega \over |\omega|}$ arises because the derivative involves an $\omega$ rather than a $|\omega|$, and the $\pm$ overall sign will not be important, so we have not computed it. 
We can now integrate out $\rho$, noticing that we can also write the quadratic term in position space, as we did for the fermionic term. 
The factor of of ${\omega \over |\omega|}$ gives an extra minus sign that is easy to miss. It  arises because we need to evaluate the linear term at $\omega$ and $-\omega$. 
In the end, we get 
\be 
 {\log Z \over N }|_{\rm bos} =  { 4 d_3^2  \over \alpha } \beta \int dt ( Q_\phi \partial_\tau Q_\phi)^2 = { 4 d_3^2  \over \alpha } \beta \int dt   Q_\phi^2 \tilde Q_\psi^2   
 \ee 
 
  We now combine this with \nref{SigCo} to obtain 
  \bea 
  {\log Z_3 \over N } &=& { 4 d_3^2\over \alpha } \beta \int dt   Q_\phi^2 (\tilde G_\psi^2 - G_\psi^2 )  
  =-{ 4 d_3^2 \over \alpha^2 \beta^2 }  \beta \int d\tau Q_\phi^2    \cr 
      & =& -  { 4 d_3^2 \over \alpha^3 }   { 1 \over (2 \pi)^2} 2 \sum_{n=1}^\infty { 1 \over n^2 } =  -{ 4 d_3^2 \over \alpha^3 } { 1 \over 12 } 
      \cr 
      & =& - { (p-2)^2 \over (p-1) 48 } { 1 \over \bar G_\phi^p}   
          \eea 
Note that the integral and the sum are the same as what appeared in \nref{V2pr}

\subsubsection{The quartic term} 

We can now consider the term that is quartic in the non-zero frequency fields. 
We get 
\bea \la{Quartic}
{\log Z_4 \over N } 
& = &   d_4 \beta \int d\tau   [ \delta G_F \delta G_\phi - 3 G_\psi^2] \delta G_\phi^2  
\\
\cr 
d_4 &\equiv & - { (p-1)(p-2)(p-3) \bar G_\phi^{p-4} 
\over 12 } 
\eea
We again add and subtract $\tilde Q_\psi^2$ to obtain 
\bea 
{\log Z_4 \over N }& =  & d_4\beta    \int d\tau   ( - \partial^2 Q_\phi Q_\phi - 3  (\partial Q_\phi)^2) Q_\phi^2   + 3 (\tilde  Q_\psi^2 - Q_\psi^2 ) Q_\phi^2 
\cr 
& =  & 3 d_4  \beta     \int d\tau     ( \tilde Q_\psi^2 - Q_\psi^2 ) Q_\phi^2
 \eea
 Where we used that we have a total derivative 
 $ \partial ( \partial Q_\phi (Q_\phi)^3 ) $ which integrates to zero. 
 For the remaining term we can also use \nref{CosSin}, as well as the leading order expression for $Q_\phi$, \nref{SolG},  which leads to  
 \bea \la{QuarticF}
 { \log Z_4 \over N } &=& - 3d_4 { 1 \over \beta^2 \alpha }    \beta \int d\tau       Q_\phi^2  
= - 3d_4  { 1 \over   \alpha^2 }  { 1 \over (2 \pi)^2 } 2 \sum_{n=1}^\infty { 1 \over n^2 }     = - { 3 d_4 \over \alpha^2 } { 1 \over 12 }  
 \cr 
 &=& {  (p-2)(p-3)   
\over 4 (p-1) }    { 1 \over   12 } { 1 \over \bar G^p_\phi}
 \eea

 Note that higher than quartic order terms in the expansion of the potential will not matter and can be neglected for our purposes.

   \subsubsection{Zero mode of $G_F$ terms } 

 When we evaluated the action in \nref{ZPartAn} we used the solution for $G_F$ from the action \nref{AcTG}. We could wonder about the effect of including higher order terms. In particular, we could consider expanding 
\be 
G_F(0) = Q_F(0) + g_F(0) ~,~~~~~~~~~Q_F(0) = { 1 \over \beta \bG^{p-1} } 
\ee 
We find that the quadratic term in the action involves 
\be 
 \propto { 1 \over Q_F(0)^2 } g_F(0)^2
 \ee 
 This is more suppressed that the other quadratic term that we included. So we do not need to consider it any further.

 \subsection{Collecting all the terms } 
 \la{Collecting}
 
We get 
\bea 
{\log Z \over N } &=& \half \log(p-1) + { \pi \over 2 \beta \sqrt{p-1} \bar G_\phi^{{ p\over 2 } -1 } } - { 1\over \bar G_\phi^p} \left[   { (p-2) \over 48 }   +{ (p-2)^2 \over (p-1) 48 } - { (p-2)(p-3) \over 48 (p-1) } \right] 
\eea
Leading to our final answer 
\be 
\boxed{ { \log Z \over N } = \half \log(p-1) + { \pi \over 2 \beta \sqrt{p-1} \bar G_\phi^{{ p\over 2 } -1 } } - { 1\over \bar G_\phi^p}  { p(p-2) \over 48 (p-1) } } \la{FinAcG}
\ee
 The factor of $(p-2)$ in the last term is clear since this term vanishes for $p=2$. In appendix \ref{GenPotA},  we make a comment on the expected answer for a more general potential. The case of $p=3$ was obtained in \cite{Lin:2013jra}. 
 The method we used here is slightly more direct. The method in \cite{Lin:2013jra} also gives the higher order corrections to the solutions. We write the results in appendix \ref{HigherOrder}. 
 
We can now maximize \nref{FinAcG} with respect to 
  $\bG$ to obtain 
\be \la{FinG}
\bG^{ {p+2 \over 2} } = \beta { p^2 \over 12 \pi \sqrt{p-1} } ~,~~~~~~~~~~~~\bG \propto \beta^{ 2 \over 2 + p} 
\ee 
and the free energy ${\cal F}$ is  
\be \la{FreeEn}
{\log Z \over N } = - \beta {\cal F} = S_0     + \hat c \, T^{ 2 p \over p+2} ~,~~~~~~~S_0 = \half \log(p-1) ~, ~~~~~\hat c \equiv 
(p+2) \left[ {  3^{ p-2}\pi^{2p } \over 2^8 p^{3p-2} (p-1)^2 } \right]^{ 1 \over p+2} 
\ee 
As a curiosity, notice that for $p=18$ we get $T^{9/5}$ which is the power that appears in BFSS. However, we will see that the physics is rather different.  
  
 It is also useful to work out the expression for $\langle \phi^2 \rangle$ which involves $G_\phi(\tau=0)$. For this purpose we should use \nref{SolG} rather than \nref{LsQ}. 
 We find that 
 \be \la{PhiSq}
 { 1 \over N } \sum_i \langle (\phi^i)^2 \rangle = \bG + { 1 \over \beta \alpha  }  \sum_{n>0} \left[ -1 + \sqrt{ 1 + {   \alpha \beta^2 \over \pi ^2 n^2 } } \right] \sim  \bG + {  1  \over \pi \sqrt{\alpha } } \log \left[ { \beta  \sqrt{ (p-1) } \bG^{p-2\over 2} 2 e^{\gamma_E-1}\over \pi }  \right]   
 \ee 
 where $\gamma_E = -\Gamma'(1)=.577..$. This is the expression plotted in Figure \ref{pl1}.

  This large $N$ analysis  breaks down for very small temperatures. We can see this by expanding \nref{FinAcG}  around the solution \nref{FinG}. Ignoring numerical constants, we get
  \be  \la{QflucG}
  \delta_2 \log Z \propto   N {  ( \Delta \bG)^2 \over \bG^{p+2} }   ~~~~ \longrightarrow ~~~~~{ \Delta \bG  \over \bG } \propto { \bG^{p  \over 2} \over \sqrt{N } } \propto { \beta^{ p \over p+2 } \over \sqrt{N} } ~,
   \ee 
   where we also indicated the typical size of the fluctuations. The solution   breaks down when the ratio $\Delta \bG/\bG$ becomes of order one, which happens at sufficiently small temperatures.

  We have seen that the leading approximation to the euclidean two point functions are given in \nref{SolG},  or \nref{LsQ} for $\tau/\beta $ not too small. We can easily continue those to Lorentzian time. In particular, the Lorentzian time version of scalar field correlator is 
  \be \la{RealTC}
  G_{\phi } = \bG - { 1 \over \sqrt{\alpha } } { t \over \beta }  + {\rm const}  ~,~~~~~~~~~~~~~~~ \alpha = (p-1) \bG^{p-2}
  \ee 
  We see that this leads to a steady decrease in the two point function. Unfortunately,  with the approximation method we used here, we stop trusting this formula when the second term becomes comparable to the first.

    In the case of the BFSS model, the gravity solution develops a ``scaling similarity'' in the appropriate range of energies \cite{Smilga:2008bt,Huijse:2011ef,Dong:2012se} which leads to the anomalous scaling of the entropy with the temperature. In this case, there is a somewhat similar phenomenon, though it is different in detail. If we look at \nref{FinAcG} and we ignore the ground state entropy which is a constant, then the $\bG$ dependent terms, the last two terms in \nref{FinAcG}, display a scaling similarity under $\beta \to \lambda \beta$ and $\bG \to \lambda^{2\over 2 + p} \bG$ where this part of the effective action gets rescaled by a factor of $\lambda^{ - {2p \over p+2}}$, which also leads to the temperature dependence in \nref{FreeEn}. Therefore, we can view this as a large $N$ theory that displays a scaling similarity similar in spirit to the one that is present in the BFSS model. One important difference is that the BFSS model does not have a ground state entropy.

   \section{Looking for a spin glass phase } 
   \la{Replicas}

In this section, we perform the quenched average to see whether the model admits a spin glass phase or a related dynamical transition. The reason we might expect a spin glass phase is the following. We have seen that $\bG$ and also $\langle \phi^2 \rangle$  become large at low temperatures. This means that the field $\phi^i$ is exploring some low valleys of the potential. So it would be natural to expect that it can get stuck in one of these valleys and not be able to thermalize efficiently by moving to neighboring valleys. 

We will explore this question by using the replica trick. For this purpose, we introduce $n$ replicas (copies) of the original system and then consider the limit $n\to 0$. For simplicity we phrase this section in the $\mathcal{N} =2$ language, with the $\mathcal{N} =4$ case being completely analogous,  as before.

When we have $n$ replicas, each field acquires an additional index $a=1,\cdots,n$. Correspondingly, since the functions of the large $N$ solution are bilocal in the fields, they acquire two indices 
\be 
G_A(t,t') \to G_A^{aa'}(t,t') ~,~~~~~~~~~~A = \phi, ~\psi,~F ;
\ee 
for example $G_{\phi } ^{aa'} (t,t')=\frac{1}{N} \phi ^{ia} (t) \phi ^{ia'} (t')$.
We should think of the $a,a'$ index as morally similar to the $t,t'$ index. It labels a position on the $n$ circles, namely the choice of which of the circles we are discussing. 
The action has the same form as we had before; the determinant term will also include a determinant on the $a,a'$ indices. The kinetic terms containing  a $\delta(t,t')$ become  $\delta(t,t')   \delta_{a,a'}$ terms. 
Similarly, the terms that arise from the interactions are changed as 
\be 
\int dt dt' G_F(t,t') G_\phi^{p-1}(t,t')  \longrightarrow \sum_{aa'} \int dt dt' G^{aa'}_F(t,t') [ G_\phi^{aa'}(t,t')]^{p-1}
\ee 	
so that also for this term, the index $a,a'$ behaves like the $t,t'$ variable.

Introducing $G$ and $\Sigma $ variables as described above, and integrating out the original fields, we get
\begin{equation}
\begin{split}
& Z_n \equiv \overline{ Z^n}= \int {\cal D}G{\cal D}\Sigma  \exp \Bigg\{ N \bigg[ \log \det \left( \delta (\tau -\tau ')\delta _{ab} \partial _{\tau } -\Sigma _{\psi}^{ ab} (\tau ,\tau ')\right) -\\
& - \frac{1}{2} \log \det \left( - \delta (\tau -\tau ')\delta _{ab}\partial _{\tau } ^2-\Sigma _{\phi}^{ ab} \right) 
- \frac{1}{2} \log \det \left( \delta _{ab} \delta (\tau -\tau ')-\Sigma _{F}^{ab} \right)  - \\
& - \sum_{ab}\int d\tau d\tau ' \left(  \Sigma _{\psi }^{ab} G_{\psi}^{ ab} +\frac{1}{2}\Sigma _{\phi}^{ ab}G_{\phi}^{ ab} +\frac{1}{2}\Sigma _{F}^{ab} G_{F}^{ab} + \frac{J^p}{2} G_{F}^{ab} (G_{\phi}^{ ab}) ^{p-1} -\frac{(p-1)J^p}{2} (G_{\psi}^{ ab} )^2 (G_{\phi}^{ ab} )^{p-2} \right) \bigg] \Bigg\}.
\end{split}
\end{equation}
 
Even when considering replica symmetry breaking configurations, we will assume that the replica matrix has diagonal components that are all equal to each other.  
When we do not write $a,b$ indices, we refer to these diagonal components. In other words, $G_A = G_{A}^{aa}$ (no sum over $a$). 
We still work in Euclidean signature, and, as is common in this type of analysis, we expect that off-diagonal components are independent of time because they involve expectation values in different replicas and we have a separate time translation symmetry on each replica; for example, for $a \neq b$, $G_{\phi } ^{ab} (\tau ,\tau ')=\frac{1}{N} \mean{\langle \phi ^{i} (\tau )\rangle \langle \phi ^{i} (\tau ')\rangle }$.\footnote{We are assuming that this time translation symmetry is not spontaneously broken by the large $N$ solution. For a somewhat similar setup where the time translation symmetry is spontaneously broken in the large $N$ solution see   \cite{Saad:2018bqo}.  }
This means that they only have zero frequency components. We also assume that the off diagonal components of $G_\psi$ are zero, since femrionic fields on each replica have zero expectation value by fermion number symmetry (this also follows from supersymmetry if it is unbroken).
Note that while one may expect the off-diagonal components of $G_{\phi } ^{ab} $ to vanish by the $\mathbb{Z} _2$ R-symmetry, it is common for this symmetry to be broken in spin glass discussions and so we will allow it to be spontaneously broken by the large $N$ solution.  
In the following,  $\bar G_A^{ab} $ stands for the Euclidean time average of $G_A^{ab} $, while $G_A^{ab} $ contain both zero and non-zero frequencies.
 
In the large $N$ limit,  we integrate out the $\Sigma_A^{ab} $ variables using their equations of motion, and we assume that $G_A(\tau ,\tau ')$ are  functions of the time difference, so that
 \bea \la{Acod}
{ \log Z_n \over N } &=&   n \sum_{\omega \not =0}\left[  - \log G_\psi + \half \log G_\phi G_F  - { \omega^2 \over 2 } G_\phi - { \half } G_F  - i \omega G_\psi  \right]  \cr  
& ~& - { n\over 2 } \beta \bar G_F + n\beta  \int_0^\beta  d\tau \left(  - \half G_F G_\phi^{p-1} + { (p-1) \over 2} G_\psi^2 G_\phi^{p-2} \right) +
\cr  
  &~&  +\half  \log \det ( \beta \bar G_\phi^{ab}  )  +\half \log \det(\beta \bar G_F^{ab}  ) - {\beta^2 \over 2 }  \sum_{a \not = b }    \bar G_F^{ab} ( \bG^{ab} )^{p-1}  .~~~~~~~~~
  \eea

  In fact, we can integrate out the zero frequency modes of $G_F^{ab}$, or $\bar G_F^{ab} $. This is useful since then the remaining off-diagonal components will only involve  $\bar G^{ab}_\phi $.
It is convenient to define 
\bea 
 M_{ab} &=&\int d\tau \, [G_{\phi } ^{ab} (\tau )]^{p-1} ~,~~~~~ {\rm or } 
 \cr 
M &\equiv& M_{aa} = \int_0^\beta  G_\phi(\tau)^{p-1} ~,~({\rm no~sum~over}~a)~,~~~~~~M_{ab} = \beta ( \bG^{ab})^{p-1} ~,~~{\rm for}~~ a\not =b
\eea
where $M$ with no indices denotes the diagonal components which are all equal. 
 The equation of motion for $\bar G_{F}^{ab}$ is
 \begin{equation}
 \beta^{-1}( \bar G_{F} ^{-1})^{ab} =\delta _{ab} +M_{ab} .
 \end{equation}
 Plugging this solution into the action we find 
 \begin{equation} \label{eq:spin_glass_action_int_out_f}
\begin{split}
& \frac{\log Z_n}{N} = n \sum _{\omega  \neq 0} \Big[ -\log G_{\psi } +\frac{1}{2} \log \left( G_{\phi }  G'_{F} \right)  - \frac{\omega ^2}{2} G_{\phi }  - \frac{1}{2} G'_F-i\omega G_{\psi } \Big] \\
& +\frac{1}{2} \log \det (\beta \bar G_{\phi }^{ab})- \frac{1}{2} \log \det (1+M_{ab} ) - \frac{n}{2}  + n \beta  \int d\tau \, \Big(- \frac{1}{2} G'_F G_{\phi } ^{p-1} +\frac{p-1}{2} G_{\psi } ^2 G_{\phi } ^{p-2} \Big) 
\end{split}
\end{equation}
where now the function $G'_F(\tau)$ is just the non-zero frequency part of $G_F$. In other words,  $G'_F(\tau) \equiv G_F(\tau) -   \bar G_F$. 
Notice that we have eliminated all the zero frequency modes   $\bar G^{ab}_F$. 

 \subsection{Replica symmetric case}
 
The analysis in the previous sections implicitly used a replica diagonal solution, with $\bG^{ab}=0$ for $a\not = b$. In order to look for a spin glass phase we should check whether there are solutions with off diagonal components.   The simplest possibility is to use a replica symmetric but non-diagonal ansatz  
\begin{equation} \label{eq:spin_glass_replica_sym_ansatz}
G_{\phi}^{ ab} = \begin{pmatrix}
G_{\phi } (\tau ) & \rsp & \cdots & \rsp \\
\rsp & G_{\phi } (\tau ) & \cdots & \rsp \\
\cdots & \cdots & \cdots & \cdots \\
\rsp & \rsp & \cdots & G_{\phi } (\tau )
\end{pmatrix} .
\end{equation}
This matrix has the eigenvalue $G_{\phi } (\tau ) + (n-1) \rsp$ with degeneracy 1 and $G_{\phi }(\tau) -\rsp$ with degeneracy $n-1$, and the same statement for its Fourier transform.  We note that in general one can get the equations of motion either by plugging in this  ansatz in the action and deriving the equations, or by using equations obtained in replica space for any replicated fields. Here we use the former approach.

We use the action \eqref{eq:spin_glass_action_int_out_f}
and plug in \eqref{eq:spin_glass_replica_sym_ansatz}
\begin{equation}
\begin{split}
& \frac{\log Z_n}{N} = n \sum _{\omega  \neq 0} \Big( -\log G_{\psi } +\frac{1}{2} \log \left( G_{\phi } G'_{F} \right)  - \frac{\omega ^2}{2} G_{\phi } - \frac{1}{2} G'_F-i\omega G_{\psi } \Big) \\
& + \frac{1}{2} \log \left(\beta  \bar G_{\phi }+(n-1)\rsp \beta \right)  + \frac{n-1}{2} \log \left( \beta \bar G_{\phi } -\rsp \beta \right)  - \frac{n}{2} \\
& - \frac{1}{2} \log \left( 1+M+(n-1) \beta \rsp^{p-1} \right) -\frac{n-1}{2} \log \left( 1+M-\beta  \rsp^{p-1} \right) \\
& -\beta n \int d\tau  \left( \frac{1}{2} G'_F G_{\phi }^{p-1} -\frac{p-1}{2} G_{\psi }^2 G_{\phi }^{p-2} \right) .
\end{split}
\end{equation}

We expand up to linear order in $n$, and then the action is proportional to $n$. This gives the $n \to 0$ limit which is the averaged $\log Z$:
%
\begin{equation} \label{eq:replica_symmetric_action}
\begin{split}
& \frac{\mean{\log Z}}{N}  = \lim_{n\to 0} { \log Z_n \over n  N} = \sum _{\omega  \neq 0} \Big( -\log G_{\psi } +\frac{1}{2} \log \left( G_{\phi } G'_{F} \right)  - \frac{\omega ^2}{2} G_{\phi } - \frac{1}{2} G'_F-i\omega G_{\psi } \Big) \\
& + \frac{1}{2} \log (\beta \bar G_{\phi }) + \frac{1}{2} \log(1-x) + \frac{1}{2} \left( \frac{1}{1-x} -2\right)  \\
& -\frac{1}{2} \log \left( 1+M- \beta\bar  G_{\phi }^{p-1} x^{p-1}  \right) -\frac{1}{2} \frac{\beta \bar G_{\phi }^{p-1}  x^{p-1} }{1+M- \beta \bar G_{\phi }^{p-1}  x^{p-1} } 
 -\beta  \int d\tau  \, \left( \frac{1}{2} G'_F G_{\phi } ^{p-1} -\frac{p-1}{2} G_{\psi } ^2 G_{\phi } ^{p-2} \right) 
\end{split}
\end{equation}
where we defined $x$  through
\begin{equation}
\rsp=\bar G_{\phi } x.
\end{equation}

For our purposes we only need the  equation for $x$  
\begin{equation} \label{eq:replica_symmetric_x_eq}
\frac{x}{(1-x)^2} =\frac{(p-1)    x^{2p-3} }{\left( r -    x^{p-1}  \right) ^2} ~,~~~~{\rm with} ~~ r \equiv { 1 + M \over Y } ~,~~~~~~Y\equiv \beta \bG^{p-1} ~,~~~~~~M = \int_0^\beta d \tau G_\phi(\tau)^{p-1} .
\end{equation}
The other equations are written in appendix \ref{FullRSB}.
 
For these solutions there is a physical constraint on the range of $x$. 
The off-diagonal elements should satisfy $q \le \bar G_{\phi } $. Denoting $\bar \phi ^a = \int d\tau \, \phi ^a (\tau )$, this inequality follows from $\langle \bar \phi ^a \bar \phi ^b \rangle  \le \langle \bar \phi ^a \bar \phi ^a\rangle $ for any fixed $a,b$. In other words, if we had the possibility of having configurations with non-zero values of $\bar \phi^a$ this quantity will be less correlated across different replicas than within the same replica.  Therefore we have the constraint\footnote{
\label{replica_symmetric_weaker_constraint_footnote}
Note that when considering fixed times for the two fields, the condition $\langle \phi (\tau )\rangle ^2  \le \langle \phi (\tau )^2\rangle $ really means that $q \le G_{\phi } (\tau =0)$ and so we get $0 \le x \le \frac{ G_{\phi } (\tau =0)}{\bar G_{\phi }} $. We expect $G_{\phi } (\tau )$ to get its maximum at $\tau =0,\beta $ (it is symmetric under $\tau  \to \beta -\tau $) and so the upper bound satisfies $G_{\phi } (\tau =0) / \bar G_{\phi }  \ge 1$. However, we see that the constraint is actually stronger than that. Note also that an $x>1$ gives an imaginary contribution to the action.}
\begin{equation} \la{xConstr}
0 \le x \le 1 .
\end{equation}

In order to have a transition to a replica symmetric solution, we should look for such a solution with $x \neq 0$ that is more dominant thermodynamically.

Let us now prove that with the  constraint \nref{xConstr} there is no solution with $x \neq 0$ at all. We will show this purely from   equation \nref{eq:replica_symmetric_x_eq}. 

First note that for $1 \le Q <P < \infty $ we have an inequality between the $P$ and $Q$ norms of a function $f$, which is $||f||_Q \le \beta ^{\frac{1}{Q} -\frac{1}{P} } ||f||_P$ where $\beta $ is the size of the interval.\footnote{This follows from Holder's inequality $||F G||_1 \le ||F||_a ||G||_b$ for $a,b \ge 1$ with $1/a+1/b=1$ by plugging in $G \to 1$, $F \to f^Q$, $a \to P/Q$ and $b \to \frac{P}{P-Q} $.} By using this inequality with $P \to p-1$ and $Q \to 1$ we conclude that 
\bea  \la{rIne}
M &= &(||G_{\phi } ||_{p-1} )^{p-1}  \ge \beta ^{2-p} \left( \int d\tau \, G_{\phi } (\tau ) \right)^{p-1}=\beta \bar G^{p-1}= Y   
~,~~~~{\rm or }
\cr 
  ~~~r& =& { 1 + M \over Y } > 1  \eea 
  where we used that  $G_{\phi } (\tau ) \ge 0$.
As a result, \eqref{eq:replica_symmetric_x_eq} simplifies to (assuming $x\not =0$)
\be 
{ 1 \over x } -1 = { 1 \over \sqrt{p-1} } \left(   { r \over x^{p-1} } -1 \right) ~,~~~~~{\rm with } ~~ r
> 1 \la{xequr}
\ee
 We see that $x=1$ is not a solution. So, assuming that $x \not =1$,  and defining $\gamma ={1 \over x}  $ we can divide by the left hand side and get the equation
\be 
 1 = { 1\over \sqrt{ p-1} } { ( r \gamma^{p-1} - 1 ) \over (\gamma -1) } > { 1\over \sqrt{ p-1} } { (  \gamma^{p-1} - 1 ) \over (\gamma -1) } =  { 1 \over \sqrt{ p-1} } \sum_{j=1}^{p-1} \gamma^{j-1} > \sqrt{ p-1} > 1 
 \ee 
 where we used that $\gamma> 1$ from \nref{xConstr},  $r>1$ from \nref{rIne}, and we used that the sum contains $p-1$ terms each bigger than one. So we conclude that we have an inconsistent equation and that \nref{eq:replica_symmetric_x_eq} has no solutions other than $x=0$.
%
%
%

We have also looked for a replica symmetric solution numerically. We have not found a solution with $0<x \le 1$. We found a solution with $x>1$ (within the range of the weaker constraint mentioned in footnote \ref{replica_symmetric_weaker_constraint_footnote}) in some range of $\beta J$, but it has an imaginary contribution to the action, and, in addition,  it is less dominant. Furthermore, it is expected to be unphysical due to the constraint \nref{xConstr}. 
 
 \subsection{1-step replica symmetry breaking}
  \def\sva{s}
  \def\uva{u}
Next we consider the  1-step replica symmetry breaking ansatz
\begin{equation}
G_{\phi } = \begin{pmatrix}
G_{\phi } (\tau ) & \cdots & \uva & \sva & \cdots  & \sva & \cdots  \\
\cdots & \cdots & \cdots & \sva & \cdots  & \sva & \cdots  \\
\uva & \cdots & G_{\phi } (\tau ) & \sva & \cdots  & \sva & \cdots  \\
\sva & \cdots & \sva & G_{\phi } (\tau ) & \cdots & \uva & \cdots \\
\sva & \cdots & \sva & \cdots & \cdots & \cdots & \cdots \\
\sva & \cdots & \sva & \uva & \cdots & G_{\phi } (\tau ) & \cdots \\
\cdots & \cdots & \cdots & \cdots & \cdots  & \cdots & \cdots 
\end{pmatrix}
\end{equation}
consisting of blocks, each of size $m$. The diagonal blocks are of the same form as the replica symmetric ansatz, and on the off-diagonal blocks we have $\sva$ everywhere. Such a matrix has the following eigenvalues 
\begin{equation}
\begin{aligned}
& G_{\phi } (\tau )-\uva \qquad && \text{with degeneracy } \frac{n}{m} (m-1),\\
& G_{\phi } (\tau )+(m-1)\uva-m\sva \qquad && \text{with degeneracy } \frac{n}{m} -1,\\
& G_{\phi } (\tau )+(m-1)\uva+(n-m)\sva \qquad && \text{with degeneracy } 1 .\\
\end{aligned}
\end{equation}

Plugging in \eqref{eq:spin_glass_action_int_out_f}, and taking the $n\to 0$ limit, with $m$ fixed, we get 
\begin{equation}
\begin{split}
& \frac{\mean{\log Z}}{N} =\sum _{\omega  \neq 0} \left( -\log G_{\psi } + \frac{1}{2} \log \left( G_{\phi } G'_F\right) -\frac{1}{2} G'_F-\frac{\omega ^2}{2} G_{\phi } -i\omega G_{\psi } \right) \\
&+\frac{1}{2} \log(\beta \bG)-\frac{1}{2} +\frac{m-1}{2m} \log(1-x)+\frac{1}{2m} \log\left( 1+(m-1)x-my\right) + \frac{1}{2} \frac{y}{1+(m-1)x-my} \\
&-\frac{m-1}{2m} \log\left( 1+M-Y x^{p-1} \right)
-\frac{1}{2m} \log\left( 1+M+(m-1) Y x^{p-1} -m Y y^{p-1} \right) \\
& - \frac{1}{2} \frac{Y y^{p-1} }{1+M+(m-1)Y x^{p-1} -m Y y^{p-1} }
 - \beta \int d\tau \left( \frac{1}{2} G'_F G_{\phi } ^{p-1} -\frac{p-1}{2} G_{\psi } ^2 G_{\phi } ^{p-2} \right) .
\\
& {\rm with } ~~~~ x \equiv { \uva \over \bG} ~,~~~~~~~y \equiv { \sva \over \bG} ~,~~~~~~~Y \equiv \beta \bG^{p-1}
\end{split}
\end{equation}
As a consistency check, note that when $x=y$, $m$ disappears from the equations and we go back to the symmetric case. Similarly,   for $m=1$ nothing depends on $x$ and reduce to \eqref{eq:replica_symmetric_action} with  $x\to y$.

It will be enough for us to look at only two  of the equations, the $y$ equation, which is 
\begin{equation} \label{eqny}
\frac{y}{\left( 1+(m-1)x-my\right) ^2} = \frac{(p-1)   y^{2p-3} }{\left( r + ( m-1) x^{p-1} - m y^{p-1} \right)^2} , ~~~~~{\rm with }~~~~~ r \equiv { 1 + M \over Y } > 1 
\end{equation}
and the $x$ equation 
\begin{equation} \label{eqnx}
\begin{split}
& -\frac{1}{1-x} +\frac{1}{1+(m-1)x-my} -\frac{my}{\left( 1+(m-1)x-my\right) ^2} +\frac{(p-1)  x^{p-2} }{ r - x^{p-1} } \\
& - \frac{(p-1)  x^{p-2} }{r + (m-1) x^{p-1} - m y^{p-1} }+ \frac{(p-1)m  y^{p-1} x^{p-2} }{(r + (m-1) x^{p-1} - m y^{p-1})^2 } =0
\end{split}
\end{equation}
We write the full set of equations in appendix \ref{Full1RSB}.

We need to make some assumptions about the range of these parameters.
 We can think about the 1-step replica solution as an extension of the replica symmetric solution into block matrices. This means that each state consists of several configurations that are close to each other with overlap $\uva$, corresponding to the previous $\rsp$, while $\sva$ is the overlap between different states \cite{mezard1987spin}. Therefore, we expect $\uva$ to be larger than $\sva$, or
\be \la{xyAssu}
x \ge y 
\ee 
In addition we have 
\be \la{xmAssu}
0 \le x, ~y \le 1 ~,~~~~~~~~~ 0 < m < 1 
\ee 
where the last one can be understood in 
the probability distribution of overlaps making sense \cite{mezard1987spin}.
Assuming $y\not =0$, we first multiply \nref{eqny} by $y$, take a square root and rearrange the terms to get 
\begin{equation}
y^{p-1} \left[ m+(1-m) \frac{x^{p-1} }{y^{p-1} } + \sqrt{p-1}\left( \frac{1}{y} -\frac{(1-m)x}{y} -m\right) \right] = r >1\end{equation}
The function on the left hand side is monotonically increasing up to $ y =x\leq 1$. 
So in the allowed range, the maximum of the left hand side is at $y=x$ where we obtain 
\be 
x^{p-1} \left[ 1 + \sqrt{p-1} \left( { 1 \over x } -1 \right) \right] = r > 1 
\ee 
But this is an inequality that follows from \nref{xequr}, which, as  we showed,    could not be obeyed. 
We conclude that the only solution of \nref{eqny} is with $y=0$. 

Now we turn to the equation for $x$, \eqref{eqnx}, which for $y=0$ and $x \neq 0$, simplifies to 
%
\be \la{EqgaSi}
 \left( \gamma  -(1-m)\right)  = \left[ { \left( r \gamma^{p-1} -1\right) \over (p-1)\left( \gamma -1\right) }\right]  \left( r \gamma^{p-1} -(1-m)\right) ~,~~~~~~~~\gamma = { 1 \over x } > 1
\ee 
We now show that the factor in square bracket is bigger than one 
\be 
 \left[ { \left( r \gamma^{p-1} -1\right) \over (p-1)\left( \gamma -1\right) }\right] >    { \left(  \gamma^{p-1} -1\right) \over (p-1)\left( \gamma -1\right) } = { 1 \over p-1} \sum_{i=1}^{p-1} \gamma^{i-1} > 1  .
\ee 
 Inserting this inequality into \nref{EqgaSi} we get 
\be 
  \left( \gamma  -(1-m)\right)  
  > \left( r \gamma^{p-1} -(1-m)\right) > \left(  \gamma^{p-1} -(1-m)\right) > \left(  \gamma   -(1-m)\right)
\ee 
which is an inconsistency. 
We conclude then that equations \nref{eqny}, \nref{eqnx} do not have non-trivial solutions in the physical range \nref{xyAssu}, \nref{xmAssu}.  This rules out a spin glass solution. We have also explored a full Parisi ansatz, in the low energy expansion, and we did not find a solution there either. 

An important point is that we have {\it not} used the equation for $m$. (It can be found in  \nref{eq:1RSB_m_eq}). As discussed in \cite{Biroli_2001} for a related model, this also implies that we are ruling out a phase similar to the dynamical phase discussed there. This is a phase where non-trivial solutions of the TAP equations \cite{TAPequations,Biroli_2001} dominate in the real time dynamics but are not relevant to the thermodynamics.

    \section{Physical interpretation } 
  
    \la{PhInterp}

\subsection{The $p=2$ model }

Here we would like to give a more physical explanation of what is going on in this model. 
   In order to guide our intuition, it will be useful to analyze the $p=2$ model from a different point of view. 
   We have briefly discussed this case in section \ref{Ptwo} where we found that, after solving all the equations except for the one for $\bG$,  the free energy is given by 
   \be \la{p2logz}
   {\log Z \over N } =  \half \log \left( { \beta \bar G_\phi \over 1 + \beta \bar G_\phi } \right) +  { \pi \over 2 \beta } 
   \ee 
   
\begin{figure}[t]
   \begin{center}
    \includegraphics[scale=.25]{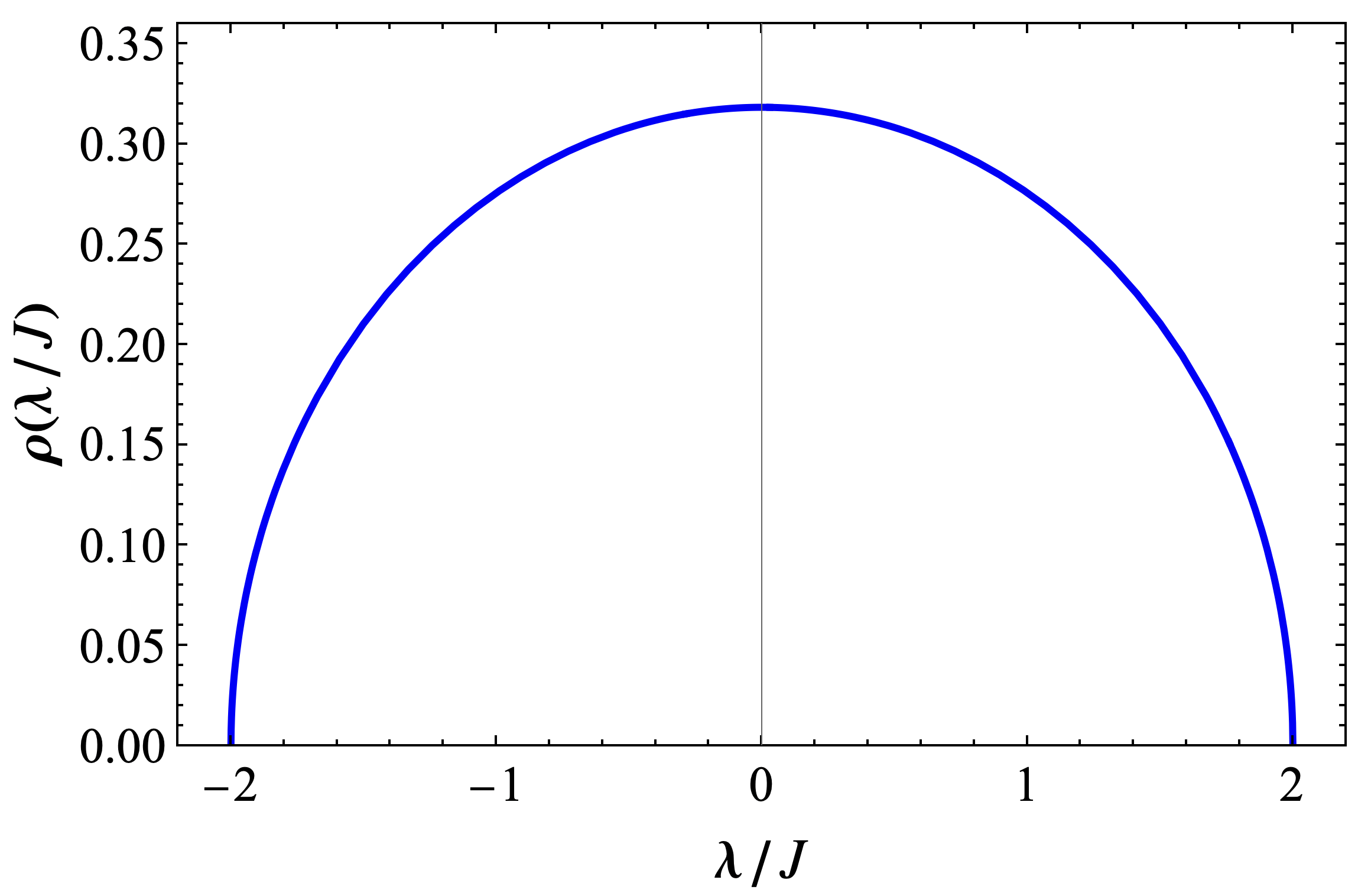}
   \end{center}
   \caption{Unit normalized eigenvalue distribution for the $p=2$ case (ordinary matrices). Note that near $\lambda \sim 0$ the distribution is essentially constant. }
   \label{eigdist}
\end{figure}

   Independently of this large $N$ analysis, we know that the quadratic model involves a random matrix $C_{ij}$ which can be diagonalized. After diagonalization we have a set of harmonic oscillators (one bosonic and one fermionic for each frequency).
   The frequencies are the eigenvalues of the matrix $C_{ij}$ and have a semicircle law distribution, see Figure \ref{eigdist}. 
   The scale of the eigenvalues $\lambda_i$ is set by the variance of the   matrix and is of order $  J$.   One important point about the eigenvalue distribution is that it is approximately independent of $\lambda$ for small $\lambda$.     This is the same as what happens for the density of states of a 1+1 dimensional field. Since the spacing is set by $1/N$, we have the same distribution as for a one dimensional field on a circle of length $L = 2 N/J$. 
   This implies that the partition function is
   \be \la{FenpTwo}
   {\log Z } =  \frac{\pi L}{4 \beta} = \frac{\pi N}{2 \beta J}
  ~,~~~~~~~{\rm for} ~~~~ \beta J \gg 1 \ee 
   which is precisely what we get from \nref{p2logz} (after restoring $J$) in the limit $\bG \to \infty$. It is important for \nref{FenpTwo} that we have an equal number of bosons and fermions so that the ground state energy vanishes. 
   
   Now we would like to explain why $\bG \to \infty$. This is something special that happens for $p=2$, but we will explain this for completeness.  By diagonalizing the matrix we find that 
   \be \la{Expphi}
   { J\over N } \sum_i \langle (\phi^{i})^{2}\rangle \sim  \frac{J}{4 \beta N} \sum_{i} \frac{1}{\lambda_{i}^{2}}
        \ee 
 where we are concentrating on the oscillators with the smallest frequencies, with  $\lambda_i \ll 1/\beta $. 
 
  We see that the infinity comes from the fact that there are some very low frequency oscillators, since the eigenvalue density is nonzero at zero frequency. This explains why $\bG=\infty$ in the large $N$ solution. If we had a specific realization of the matrices, then we expect that the lowest frequency would be of order $\lambda_{min} \sim 1/N$. This would make \nref{Expphi}  finite. However, if we average this over disorder, then we will still get infinity, even at finite $N$,  because the integral over the eigenvalues is not suppressed near zero and we get a divergence from the factor of $1/\lambda_i^2$. This implies that the replica method should give infinity even at finite $N$. 
  
  One lesson from this discussion is that small eigenvalues of the matrix $C_{ij}$ are important because they represent nearly flat directions in the potential. 
  
  \subsection{The $p>2$ case } 
  
    We now consider the case $p>2$. We now have a multi-index random tensor. However, the lesson from the $p=2$ case is that nearly flat directions in the potential are important. Since the potential is homogeneous, we should be exploring the directions along which the potential is very small. We can explore this by looking for the analog of the eigenvectors
        minimizing the potential
\be
	V = \half \sum_{i} \lb \partial_{i}W(\phi) \rb^{2}, \quad  \quad W = C_{i_{1}\cdots i_{p}}\phi^{i_{1}}\cdots \phi^{i_{p}}
\ee
subject to the constraint
\begin{align}\label{Gconst}
	\frac{1}{N} \phi_{i}\phi^{i} = \bG
\end{align}
which we impose by a Lagrange multiplier. 

This leads to an equation of the form
\begin{align}\label{teig}
	C_{i_{1}i_{2}\cdots i_{p}}\hat \phi^{i_{2}}\cdots \hat \phi^{i_{p}} = \lambda \hat \phi_{i_{1}}, \quad \quad \frac{1}{N} \hat \phi_{i} \hat \phi^{i} = 1
\end{align}
Here $\lambda$ is a number related to the Lagrange multiplier. Solutions $( \hat \phi_{i}, \lambda)$ to this equation are called normalized eigenvectors and eigenvalues of the tensor $C$ \cite{Cartwright_2013,Qi:2005pmd}.\footnote{In  \cite{Cartwright_2013,Qi:2005pmd} they use a different normalization for the eigenvector, so the eigenvalues differ by a simple power of $N$.} $C$ is a real, symmetric tensor, which in general has complex eigenvectors and eigenvalues. If an eigenvector of $C$ is real, then the corresponding eigenvalue is also real. 

We can ask how many   solutions to \nref{teig} there are. As in the computation of the index, we expect exponentially many complex solutions, in this case $(p-1)^{N}-1$ solutions for large $N$ \cite{Garcia_Li}. Out of these, the typical number of real solutions, $\overline{ \Omega_{p }}$,  was shown to scale as \cite{auffinger2011random} 
\begin{align} \la{TotNSo}
	\lim_{N\to \infty} \frac{1}{N} \log \overline{\Omega_{p}}  = \frac{1}{2}\log (p-1)
\end{align}
This agrees with the ground state entropy in the large $N$ effective action \nref{FinAcG}, and the reason will become clearer below.  

When $p=2$, \nref{teig} reduces to an ordinary matrix eigenvalue equation. In this case, the eigenvalue distribution in the $N \to \infty$ limit is well known to be given by the Wigner semicircle law.
 
 It is interesting to ask about the eigenvalue distribution of tensors. This question was investigated in \cite{auffinger2011random, Sasakura:2022iqd,
 Sasakura:2022axo,Evnin:2020ddw}, see also    \cite{Gurau:2020ehg}. 
    These investigations have shown that the distribution is nonvanishing around $\lambda \sim 0$. In particular, they suggest that the eigenvalue distribution is Gaussian around the origin with a  width scaling like $\frac{1}{\sqrt{N}}$.\footnote{They also suggest that the maximum eigenvalue is independent of $N$ at large $N$.}  Moreover, the density of eigenvalues contains an exponential in $N$ prefactor coming from \nref{TotNSo}.   In other words, these distributions are smooth, nonzero functions around $\lambda \sim 0$ that are multiplied by an exponentially large factor  of order $e^{S_0}$ or $(p-1)^{N \over 2} $. 
  This means that there are an exponentially large number of eigenvalues near the origin. This leads to an exponentially large number of nearly flat directions. This exponential number gives rise to the ``ground state entropy'' $S_0$ we had in \nref{FinAcG}.

 Note that our computation of the index suggests that there will be no zero energy states, at least for odd $p$ and odd $N$. However, in the large $N$ analysis, as well as in this eigenvalue analysis, we find an exponentially large number of very low energy states. We interpret this number as giving rise to $S_0$. 

We should now discuss what happens as the fields move along one of these very flat directions. 
Let us say that we have a field configuration of the form 
\be 
\phi^i = \sqrt{\bG} \hat \phi^i_\alpha  + \delta \phi^i 
\ee 
where $\hat \phi^i_\alpha $ is an eigenvector with a low eigenvalue and $\delta \phi^i$ are fluctuations around this direction. 
We   want to understand what potential these fluctuations see. To first order we get nothing since the $\hat \phi^i_\alpha$ are extrema.\footnote{They are not extrema in the radial direction, but this is only one out of $N$ directions, which is an effect subleading in $1/N$.} So we need to expand to second order  
\be 
 W = \hat C_{ij} \delta \phi^i \delta \phi^j  ~,~~~~~~~~ \hat C_{ij} \equiv {p(p-1) \over 2} C_{ijk_1 \cdots k_{p-2} } 
 \hat \phi_{\alpha}^{k_1} \cdots  \hat \phi_{\alpha}^{k_{p-2}} (\bG)^{p-2 \over 2 } 
\ee 
Now, the important point is that we get a matrix $\hat C_{ij}$ which is effectively random and its variance can be estimated as 
\be 
 \mean{ \hat C_{ij}^2 } = \frac{p^{2}(p-1)^{2}}{4}{  \bG^{p-2}   }  \mean{  ( C_{ijk_1 \cdots k_{p-2} } 
 \hat \phi_{\alpha}^{k_1} \cdots  \hat \phi_{\alpha}^{k_{p-2}} )^2 } \propto \frac{(p-1)J^{p} \bar G_{\phi}^{p-2}}{4 N}
 \ee 
 where we have used the average values of the $C$'s in the second term.

 This is non-trivial since $\hat \phi_\alpha$ depends on the specifics of the $C$'s, and indeed if we had $p-1$ insertions of the eigenvector, we could use the eigenvalue equation. Here we are assuming that if we have only $p-2$ insertions, we will get an essentially random answer which is given by the average value. We do not have a very good justification for this estimate other than it reproduces the answer from the large $N$ equations, as we will see. 
 
 We then conclude that the $\delta \phi^i$ fluctuations see a random matrix whose overall energy scale is set by $\bG^{p-2\over 2}$. Then we can compute their contributions to the free energy using the results we obtained above for the $p=2$ model \nref{FenpTwo}. This reproduces the first term that we found in the effective action \nref{ZPartAn}.
 So the conclusion is that the result in \nref{ZPartAn} is coming from small fluctuations around the flat valleys, and these small fluctuations look like harmonic oscillator modes with random masses.
 Because the energy scale of the harmonic oscillators is set by $\bG$, the final free energy does depend on $\bG$ and it wants to push $\bG$ to lower values. The reason is that as we lower the value of $\bG$, for a fixed $\beta$, we are lowering the frequencies of the harmonic oscillators. This increases the entropy more than it contributes to $\beta E$. 
 
 Note that it is important that the ground state energies of the bosons and the fermions are cancelling due to supersymmetry. If that were not to happen, then we would get an additional energy contribution which would contribute as  $\pm \beta \bG^{p-2\over 2} $ (note the positive power) where the $\pm$ depends on whether we have more fermions or bosons. As an example, we can consider a truncation of our model where we remove the fermions, leaving the bosons, with the same potential for the bosons. In this case, the discussion of the flat directions of the potential is initially similar, but when we get to the harmonic oscillators, we find that there is a positive contribution to the energy of the form 
 $ \bG^{p-2\over 2}$, which pushes $\bG$ to small values, especially at low temperatures. We discuss the large $N$ solution of this model in appendix \ref{BosonicModel}.

 We can now wonder about the origin of the $1/\bG^p$ term in the effective action \nref{FinAcG}. 
These terms appear to arise from the interactions between the harmonic oscillators. For example, let us consider the term that arises from the quartic term \nref{Quartic}. For simplicity, 
  assume that $p=4$. Then we would have the microscopic term 
 \be 
  C_{ijkl} ( \delta F^i \delta \phi^j \delta \phi^k \delta \phi^l  + \psi_1^i \psi_2^j \delta \phi^k \delta \phi^l  ) 
  \ee 
    Taking the disorder average and the expectation value of the square of this term we obtain the quartic term considered in \nref{Quartic}, which leads to the $1/\bG^p$ term \nref{QuarticF}. 
 We conclude that the $1/\bG^p$ terms come from the interactions between the harmonic oscillators. This is not surprising since that is precisely how we got these terms in the effective action method of section \ref{LowEn}. 
 Note that these higher order terms are sensitive to the amplitude of the oscillations. So the higher the effective mass scale of the oscillators, the smaller the term. So these interactions produce a term that becomes large for   smaller values of $\bG$, where the amplitudes get bigger. The sign of this term is such that it pushes $\bG$ to larger values. 
 
 The conclusion is that there are fairly flat valleys in the potential. The quadratic fluctuations around these valleys induce some dependence on the value of the typical scale of the vacuum expectation value of the $\phi$s. In addition, further interactions introduce additional terms that stabilize the scale of $\phi$, or $\bG$, at a finite value. This value depends on the temperature. 
 
 The effective potential for $\bG$ becomes smaller and shallower for smaller temperatures. 
 We can estimate the temperature at which the large $N$ approximation breaks down as we discussed around \nref{QflucG}. We have not explored what happens at even lower temperatures.

  This picture, with the various valleys, would suggest that the system picks out a direction and stays there, especially at low temperatures where $\bG$ is large. For this reason, it is a  bit surprising  that the replica analysis did not display any   solutions with off diagonal components. In fact, as originally suggested by \cite{Anninos:2016szt}, we also expected to find a spin glass phase. 
  However, the detailed analysis of section \ref{Replicas} showed that this is not the case.  It seems to suggest that the fields can   easily be taken  from one valley to another by thermal or quantum fluctuations,  so that they are not stuck in any single one, or that all these valleys are essentially connected to each other. 
  
  Note that the absence of one replica symmetry breaking solution was shown imposing just a subset of the equations. In particular, the equation that governs the sub-block size $m$ was not imposed. According to \cite{Biroli_2001,cugliandolo2023recent} this is saying that we do not even have a more subtle phase, where the dynamics gets stuck in particular directions.  Notice that the real time correlator computed by the replica diagonal correlator   in \nref{RealTC} shows   that $\langle \phi^i(t) \phi^i(0) \rangle $ decreases below $\bG$ at sufficiently large real times. 
  
  A peculiar property of this model is that supersymmetry implies that there is a minimum energy at zero and that all these valleys have similarly small energies. Perhaps this is important for the absence of the spin glass-like phases. 
  
  We could consider small variants of the model, where we impose a supersymmetric version of the spherical constraint $\sum_i \phi_i^2 = $constant. When this constant is large, we expect a similar physical picture, for the reasons explained above. But it would be interesting to study it in detail. 
  
  There are other variants, were we could set the $C$s to zero when any of the indices are equal to each other. The large $N$ limit would be the same, but such a theory would have some classically exact flat directions.

\subsection{Comparison with other models} 
  \la{OtherModels}
  
  \subsubsection{The $p$-spin model } 
  
 A related model is the $p$-spin model. It consists of $N$ scalar fields with a potential 
  \be 
   V = C_{i_1 \cdots i_p } \phi^{i_1} \cdots \phi^{i_p} ~,~~~~~~~  \sum_i (\phi^i)^2 = N g
   \ee 
   and they are subject to a spherical constraint. This spherical constraint is somewhat similar to the situation we had in our model when we were fixing $\bG$. Here the difference is that the local minima of the potential can have a range of values of the potential energy.  In terms of the eigenvalues of the tensor $C$ we would look for eigenvalues near the edge of the distribution, see \cite{Evnin:2020ddw}. Perhaps the fact that we are trying to minimize $V$ would give a larger propensity for the fields $\phi^i$ to get stuck around some direction, which is what seems to happen in this model \cite{Anous:2021eqj}. This model has a one step replica symmetry breaking solution. The fact that it is one step replica symmetry implies that there are only two  possible overlaps between different directions that are approximate minima of the potential. One possible overlap is zero, which is what we expect when we have two different random directions on a high dimensional sphere. The other is a non-zero overlap which is what we have when we choose one direction and let the field explore the part of the space that is accessible within that region \cite{cugliandolo2023recent}.  
 
   In the $p$-spin model the entropy vanishes at low temperatures at leading order in $N$ \cite{Anous:2021eqj}.  This suggests that the number of directions that the system can explore is not exponentially large, in contrast with what we found in the supersymmetric model in this paper.

\subsubsection{Bosonic model that results from removing the fermions of the SUSY model} 
\la{TruncMod}

Another model that we can consider is a truncation of the supersymmetric model where we remove the fermions. This model contains just the bosons $\phi^i$ and a potential which is of the form $V \sim (\partial_i W)^2 $ which has a minimum at $\phi^i=0$. 

The large $N$ equations of this model are the same as in the supersymmetric model, except that we set $G_\psi=0$. In particular, it is also convenient to introduce the auxiliary fields $F^i$ as well as the function $G_F$. 
Due to the absence of the fermions,  the low temperature physics is very different. In particular, let us consider what happens when  $\phi^i$ wants to develop a large vacuum expectation value $\bar \phi^i$, by exploring the nearly flat directions of the potential. As we discussed above, in that region, the small fluctuations of $\phi$ around  $\bar \phi^i$  is described by a collection of harmonic oscillators coming from diagonalizing the mass matrix of the quadratic fluctuations. These oscillators have a ground state energy that is positive and proportional to a power of $ (\bar \phi^i)^2 $. This would then drive $\bar \phi^i$ to small values. This did not happen in the supersymmetric model because there was a cancellation between the ground state energy of the fermions versus that of the bosons. This suggests that the low temperature physics of this model might be gapped. Indeed, this is what we find by a more detailed analysis described in appendix \ref{BosonicModel}. In addition, we show there that $J \langle \phi_i^2 \rangle $ becomes a constant at low temperatures, as expected from this argument.

   \section{Conclusions}

  We have discussed here some ${\cal N}=2,4$ supersymmetric models with a special low energy behavior. 
  We studied the thermodynamics of these models by looking at the large $N$ equations  both numerically and analytically. The low energy analytic approximation gives us some insight into the low energy physics of the problem. The absence of replica symmetry breaking solutions suggests that these are the right equations describing the low energy limit. 
    We presented a physical picture where the scalar fields get a large vacuum expectation value exploring very shallow directions in the potential. The quadratic fluctuations around this large value give rise to an effective potential that determines the vacuum expectation value of the field and gives rise to the fractional power of temperature in the low energy thermodynamics. The lack of a spin glass phase suggests that the various shallow valleys are connected to each other.  
        
    The fact that the main dynamical effects come from the quadratic fluctuations suggests that this model will not have a maximal Lyapunov exponent. In principle, one could  check this directly by computing the four point function, but we have not done it.     
    
   While the initial motivation to look at these models was based on some similarities with BFSS, the physics looks fairly different. These models have a ground state entropy, while BFSS does not. In addition, BFSS is expected to be maximally chaotic in the regime that it is described by gravity, while these models are not. At zero temperature the expectation value of the scalars diverges in these models, while in BFSS we expect that $\langle Tr[X^2] \rangle$ is finite \cite{Lin:2014wka}.

\subsection*{Acknowledgments}

We would like to thank T.~Anous, L.~Cugliandolo, D.~Ghim, F.~Haehl, D.~Huse, Z.~Komargodski, J.~Kurchan, B.~Swingle, M.~Winer,  and the participants of the Summer Simons workshop at Stony Brook for useful comments and suggestions.   

J.M. is supported in part by U.S. Department of Energy grant DE-SC0009988.
VN is supported in part by the US National Science Foundation under Grant No.\ PHY-2209997.


\appendix

\section{Bosonic model that results from removing the fermions } 
\la{BosonicModel}
In this section we discuss a variant of the model in which we remove the fermions. Clearly this model is not supersymmetric. However, to help us understand the physics of the original model, it is interesting to ask whether the growth of $\bar G_{\phi}$ at low temperatures persists when the fermions are absent. 

This model contains dynamical bosons $\phi^{i}$ and  auxiliary fields $F^{i}$, $i = 1,\cdots,N$. For $p=3$, the action   is
\begin{align}
	S = \int d\tau \lb \frac{\dot \phi^{2}}{2}+\frac{F^{2}}{2}+3iC_{ijk}F^{i}\phi^{j}\phi^{k}\rb
\end{align}
As before, the $F^{i}$ can be integrated out to produce a potential $V \sim \sum_{i} \lb \partial_{\phi_{i}}W \rb^{2}$, where $W(\phi) = C_{ijk}\phi^{i}\phi^{j}\phi^{k}$. To obtain the action for general $p$, we replace $C_{ijk}\phi^{i}\phi^{j}\phi^{k}\rightarrow C_{i_{1},\cdots,i_{p}}\phi^{i_{1}}\cdots\phi^{i_{p}}$.

To study the large $N$ solution, we perform the Gaussian integral over disorder and write an effective action in the bilocal variables $G, \Sigma$ at leading order in $N$. These are the same manipulations as described in section \ref{lgNeq}. From the effective action we derive the equations of motion. They can also be obtained simply by setting $G_{\psi}(\tau) = \Sigma_{\psi}(\tau) = 0$ in the Schwinger-Dyson equations \nref{eq:SD_eqs_Sigma}, \nref{eq:SD_eqs_G_general_q}. The finite-temperature Euclidean equations are
\begin{align}
	G_{\phi }(\omega) &= \frac{1}{\omega^{2}-\Sigma_{\phi}(\omega)}, \quad \quad \Sigma_{\phi }(\tau) = -(p-1)J^{p}G_{\phi}^{p-2}(\tau)G_{F}(\tau)\label{sdb1}\\
	G_{F}(\omega) &= \frac{1}{1-\Sigma_{F}(\omega)}, \quad \quad \Sigma_{F}(\tau) = -J^{p}G_{ \phi}^{p-1}(\tau)\label{sdb2}
\end{align}

\subsection{Numerical solution}

Here we discuss a numerical solution of  \nref{sdb1}, \nref{sdb2} in the $p=3$ model. We use the same method of weighted iterations as discussed in section \ref{numsec}. We find that at low temperatures, $\frac{J\langle \phi^{2}\rangle}{N} $ decreases monotonically, approaching a constant value. 
This is consistent with our expectation based on the physical picture described in section \ref{TruncMod}. As before, we have a large number of shallow directions of the potential corresponding to solutions of a tensor eigenvalue equation with near-zero eigenvalues. When $\phi^{i}$ explores one of these flat directions, bosonic fluctuations in the orthogonal directions acquire large masses that contribute to the effective potential.
In this case, there are no states in the fermionic sector to cancel the ground state energies of the bosons. So, the free energy contribution from these massive modes prevents $J \langle \phi^{2} \rangle /N$ from becoming large. In fact, we will argue that the model is gapped at low temperatures.

\begin{center}
\includegraphics[height=8cm]{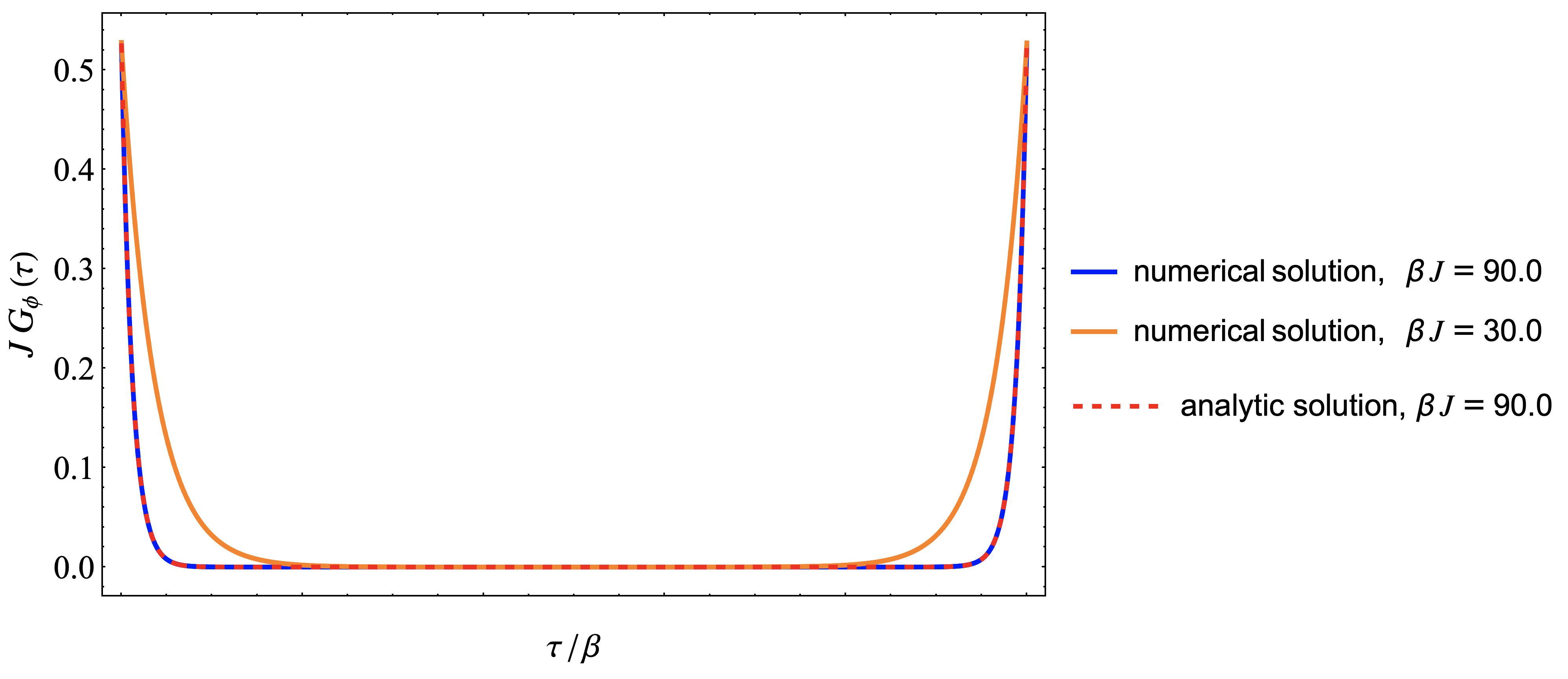}
\includegraphics[height=8cm]{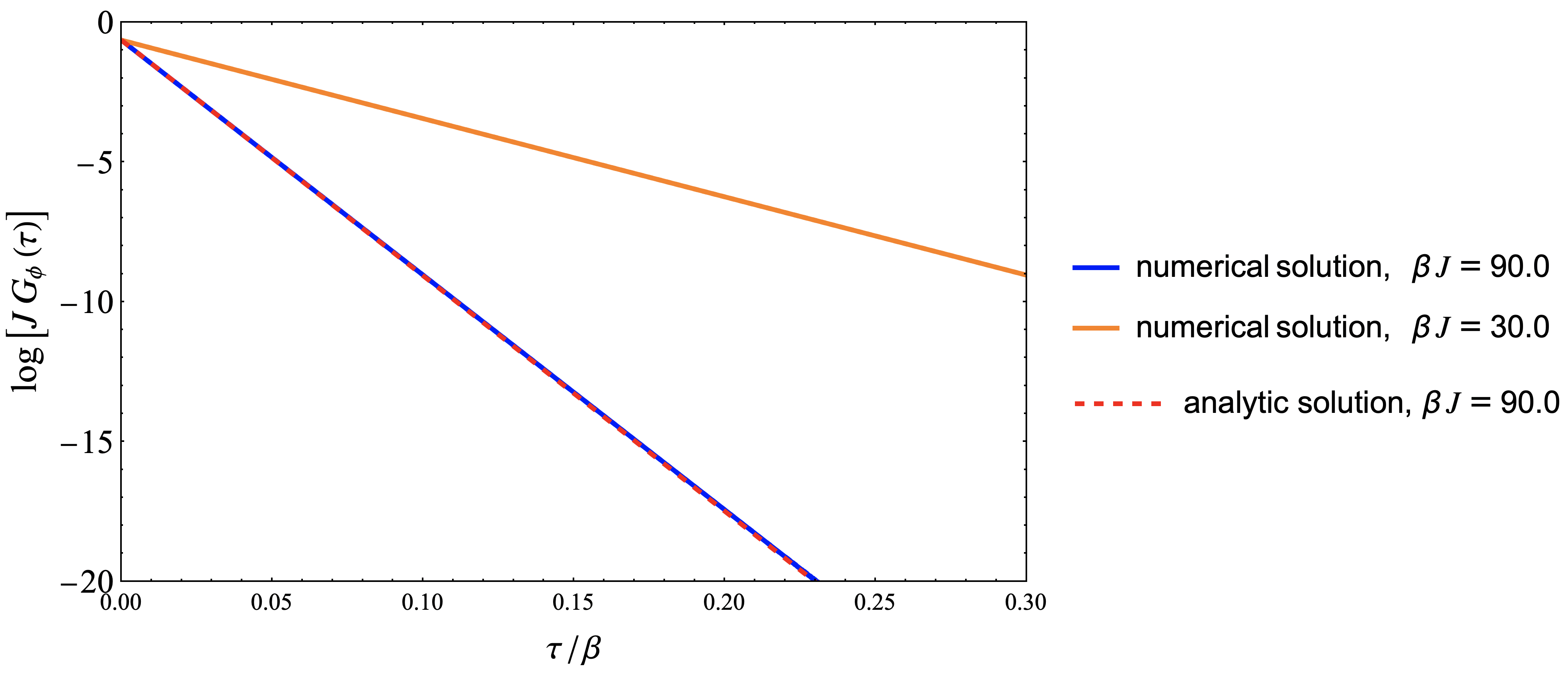}
\includegraphics[height=7.7cm]{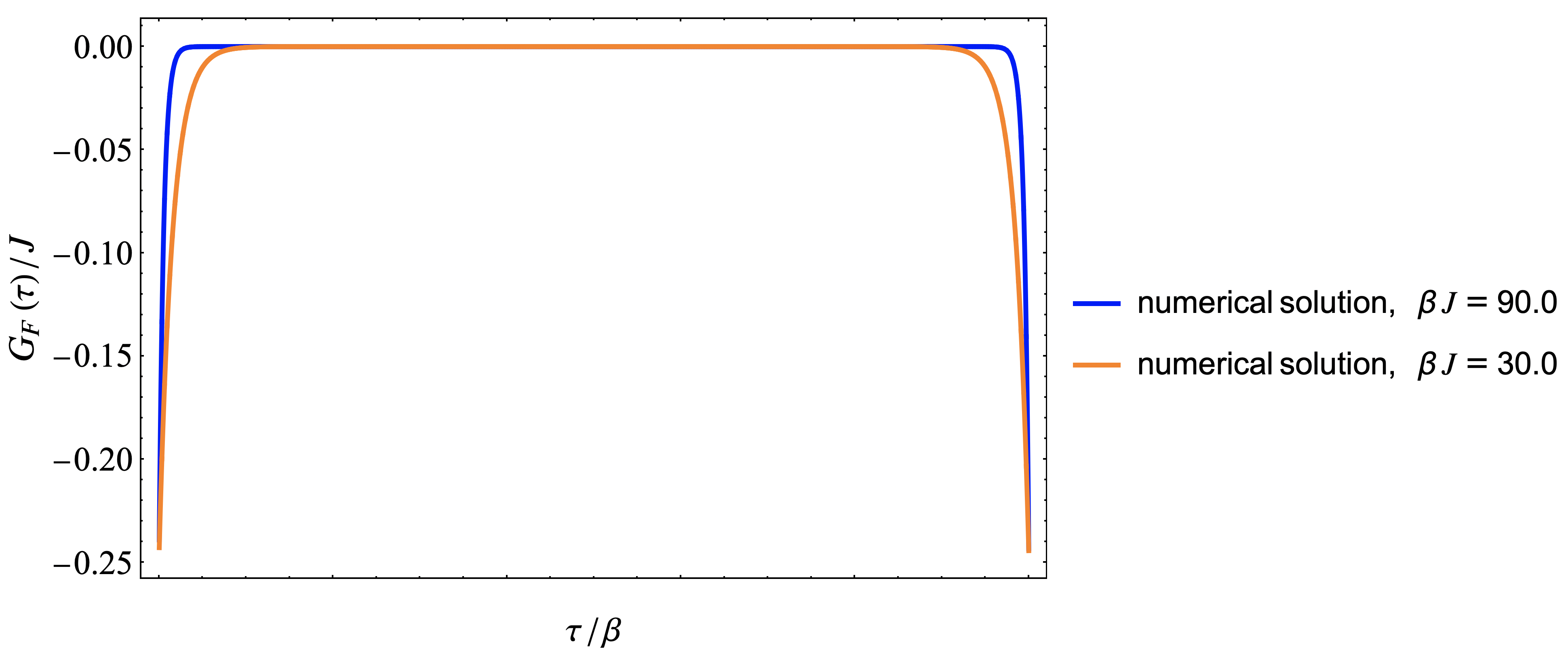}
\includegraphics[height=7.7cm]{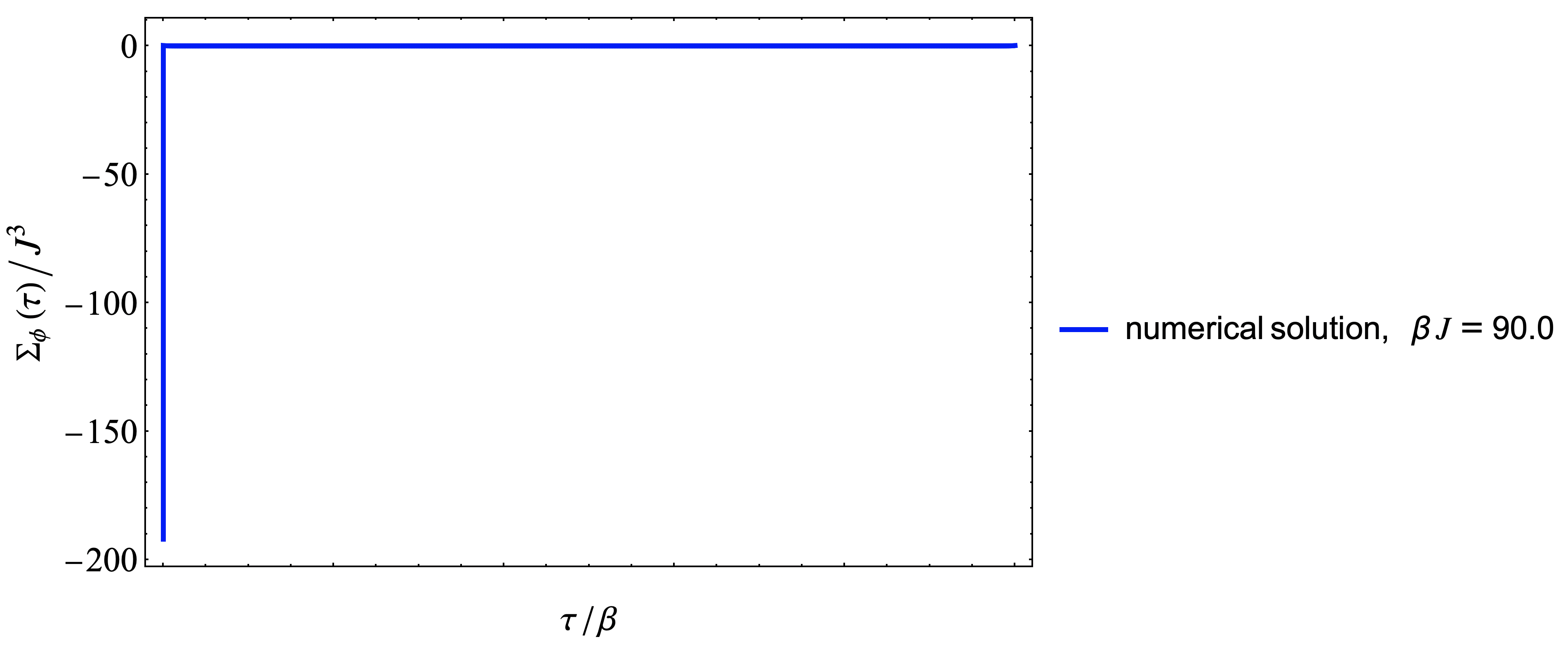}
\captionof{figure}{Numerical solutions for $J G_{\phi}(\tau)$, $G_{F}(\tau)/J$, and $\Sigma_{\phi}(\tau)/J^{3}$ in the $p=3$ bosonic model at various $\beta J$. In the plots of $J G_{\phi}$, the low-temperature analytic prediction \nref{bosGsol} is also shown. This prediction is derived in section \nref{anbos}. The fitted constants are $c_{1} = 0.936$ and $c_{2} = 0.527$ with standard errors of order $10^{-5}$. We see from the linearity of $\log[J G_{\phi}(\tau)]$ that the correlator decays exponentially, indicating a mass gap. At low temperatures, $\Sigma_{\phi}(\tau)/J^{3}$ is concentrated near $\tau = 0$, which will motivate the analysis in section \nref{anbos}.} \label{bosplts}
\end{center}

\begin{center}
\includegraphics[height=8cm]{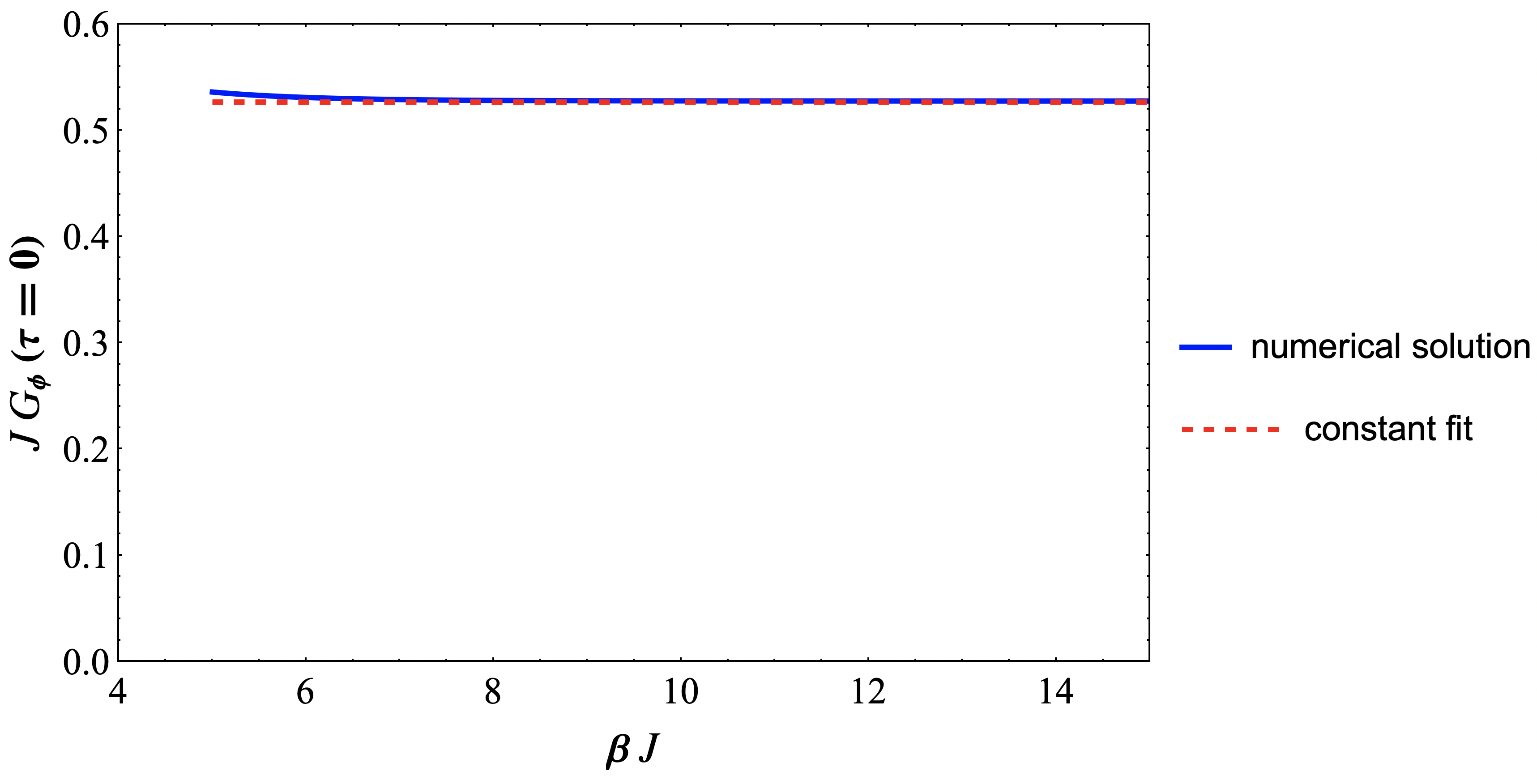}
\captionof{figure}{Numerical solution for $J G_{\phi}(\tau = 0)$ in the $p = 3$ bosonic model. We also plot a constant fit to the function at large $\beta J$, $f(\beta J) = 0.527$. } 
\label{phvevb}
\end{center}

\subsection{Analytic solution for $\beta J \gg 1$}\label{anbos}

In this section, we discuss an analytic approximation for the low temperature solution.

The Schwinger-Dyson equations can be expressed in the time domain as
\begin{align}
	-&\partial_{\tau}^{2}\tilde G_{\phi}(\tau)-[ \Sigma_{\phi}*\tilde G_{\phi}](\tau) = J \delta(\tau)\label{Gphbos}\\
		&G_{F}(\tau) - [\Sigma_{F}*G_{F}](\tau) = \delta(\tau) \label{Gfbos}\\
		&\Sigma_{\phi}(\tau) = -(p-1)J^{2}\tilde G_{\phi}(\tau)^{p-2}G_{F}(\tau)\label{sigmaphbos}\\
	&\Sigma_{F}(\tau) = -J \tilde G_{\phi}(\tau)^{p-1}\label{sigmafbos}
\end{align}
where $\tilde G_{\phi}(\tau) = J G_{\phi}(\tau)$ and $*$ denotes the convolution
\begin{align}
	[ \Sigma_{A} * G_{A}] (\tau) = \frac{1}{\beta}\sum_{\omega} e^{-i \omega \tau}\Sigma_{A}(\omega) G_{A}(\omega)
\end{align}
From the numerical solution, we notice that  $\Sigma_{\phi}(\tau)/J^{3}$ is concentrated near $\tau = 0$ at low temperatures, see Figure \ref{bosplts}. This suggests that the convolution in \nref{Gphbos} will be localized in time and simply average the function $G_\phi(\tau)$ over a small range in times, which to leading order can be viewed as multiplying the function by a constant, which we can call $m^2$. Then \nref{Gphbos} becomes 
\begin{align}\label{meqn}
	-\partial^{2}_{\tau} \tilde G_{\phi} + m^{2}\tilde G_{\phi}(\tau) =J \delta(\tau)
\end{align}
This is the equation for a free massive field in Euclidean signature. It admits the solution
\begin{align}\label{bosGsol}
J G_{\phi}(\tau) =\frac{ c_{2} \cosh \lb c_{1} J \lp \tau - \frac{\beta}{2} \rp\rb }{   \sinh \lb c_{1} J \frac{\beta}{2} \rb } \quad \text{with} \quad c_{1} = \frac{m}{J}, ~~c_{2} = \frac{J}{2m}
\end{align}
where we have used the boundary condition $\partial_{\tau}\tilde G_{\phi}(0) = -\frac{J}{2}$ to determine the prefactor.

The functional form of this solution agrees well with the numerical data. Fitting \nref{bosGsol} to the numerical solution, we find the best fit constants $c_{1} = 0.936$ and $c_{2} = 0.527$, with standard errors of order $10^{-5}$. Moreover, the best fit constants are independent of $\beta J$ for $\beta J$ sufficiently large. The analytic and numerical solutions for $J G_{\phi}(\tau)$ are plotted in Figure \ref{bosplts}. 

Furthermore, this analytic solution predicts that  $J G_{\phi}(\tau = 0) $ approaches a constant at low temperatures,
\begin{align}\label{vevcon}
	J G_{\phi}(\tau = 0) \rightarrow c_{2} \quad \text{ as } \quad \beta J \to \infty
\end{align}
Indeed, Figure \ref{phvevb} shows that $J G_{\phi}(\tau = 0)$ is approaching a constant value of about $0.527$, which coincides with the numerical result for $c_{2}$. 

The conclusion of this analysis is that, when the fermions are removed, $\frac{J \langle \phi^{2} \rangle}{N}$ does not grow at low temperatures but decreases monotonically, approaching a constant value. In this regime, the $\phi$ correlator is exponentially decaying, indicating the presence of a mass gap. 

Notably, $m$ is a free parameter in our analytic solution. We will now discuss one way to get an analytic estimate for the value of $m$.
 
Substituting $\Sigma_{F}$ into the equation for $G_{F}$ \nref{Gfbos},
\begin{align}
	G_{F}(\tau) = \delta (\tau) - J [\tilde G_{\phi}^{p-1} * G_{F}](\tau)
\end{align}
$\tilde G_{\phi}(\tau)$ is becoming small at low temperatures, at least at intermediate times. This motivates us to approximate $G_{F}(\tau)$ as a delta function. Using \nref{sigmaphbos}, the equation for $\tilde G_{\phi}(\tau)$ \nref{Gphbos} becomes
\begin{align}
	-&\partial_{\tau}^{2}\tilde G_{\phi}(\tau)+ m^{2}\tilde G_{\phi}(\tau)  = J \delta(\tau), \quad \quad m^{2} = (p-1)J^{2}\tilde G_{\phi}(\tau = 0)^{p-2}
\end{align}
Evaluating our solution \nref{bosGsol} at $\tau = 0$ then gives an equation for $m$,
\begin{align}
	\lp \frac{m^{2}}{(p-1)J^{2}}\rp^{\frac{1}{p-2}} = \frac{J}{2m \tanh (m \beta/2) }
\end{align}
This equation can be solved in the low temperature limit:
\begin{align}\label{m}
\frac{m}{J} = (p-1)^{\frac{1}{p}}2^{\frac{2-p}{p}}
\end{align}
Therefore we estimate $c_{1}$ and $c_{2}$ to be given by
\begin{align}
	\bar c_{1} = \frac{1}{2 \bar c_{2}} = (p-1)^{\frac{1}{p}}2^{\frac{2-p}{p}} 
\end{align}
where we use a bar to denote predicted values. For $p = 3$, $\bar c_{1} = 1$ and $\bar c_{2} = \frac{1}{2}$, which is pretty close to the values we found numerically.

  \section{ Possible addition of a $G_{\phi F}$ correlator to the large $N$ equations } \label{sec:Phi_F}
In the main text we have set $G_{\phi F} =0$. In this appendix we show that the solution is stable around that configuration. 

Restoring these fields, 
and plugging them into the effective action we get the following terms 
\begin{equation}
\begin{split}
\frac{\log Z}{N} & \supset -\frac{1}{2} \log \det \begin{pmatrix} -\partial _{\tau } ^2 - \Sigma _{\phi \phi } &-\Sigma _{\phi F} \\ -\Sigma _{F\phi } & \delta (\tau -\tau ')-\Sigma _{FF}  \end{pmatrix}
+ 
\\
& + \int d\tau d\tau ' \Bigg( -\frac{1}{2} \Sigma _{\phi \phi } G_{\phi \phi }
-\frac{1}{2}\Sigma _{FF} G_{FF} -\Sigma _{\phi F} G_{\phi F} -\frac{1}{2} J^p G_{FF} G_{\phi \phi } ^{p-1} \\
& 
- \frac{p-1}{2} J^p G_{\phi F} G_{F\phi } G_{\phi \phi } ^{p-2} \Bigg) .
\end{split}
\end{equation}
where we ignored the terms involving $G_\psi$ and $\Sigma_\psi$ since they do not play a role in our analysis below. 

Now it is possible to integrate out all the $\Sigma $'s. In general, if we have an expression of the form $ \pm \frac{1}{2} \log \det (-\Sigma _{ab} +{\cal M}_{ab} )- \frac{1}{2}\sum _{ab} \Sigma _{ab } G_{ab} $, the equations of motion give us $(-\Sigma +{\cal M})_{ab} =\mp G^{-1} _{ba} $. Plugging it back in the original expression, we find, up to a constant that has no significance to us, $ \mp \frac{1}{2} \log \det G - \frac{1}{2}  \sum _{ab} {\cal M}_{ab} G_{ab} $. 
 As a result
\begin{equation}
\begin{split}
\frac{\log Z}{N} &\supset  
 \frac{1}{2} \log \det \begin{pmatrix}G_{\phi \phi } & G_{\phi F} \\ G_{F\phi } & G_{FF} \end{pmatrix}+
\sum _{\omega } \left[ 
-\frac{\omega ^2}{2} G_{\phi \phi } -\frac{1}{2} G_{FF} \right] +\\
&+ \int d\tau d\tau ' \Bigg( -\frac{1}{2} J^p G_{FF} G_{\phi \phi } ^{p-1}
- \frac{p-1}{2} J^p G_{\phi F} G_{F\phi } G_{\phi \phi } ^{p-2} \Bigg) .
\end{split}
\end{equation}

We now expand in  $G_{\phi F} $ up to quadratic order to obtain 
\begin{equation}
\frac{\log Z}{N} \supset-\frac{1}{2} \sum _{\omega } \frac{G_{\phi F} (\omega )^2}{G_{\phi \phi } G_{FF} (\omega )} - \frac{p-1}{2} J^p \beta \int d\tau \, G_{\phi \phi } ^{p-2} G_{\phi F} (\tau )^2.
\end{equation}
Now we note that $G_{\phi \phi } (\tau ) \ge 0$. In addition, $G_{\phi \phi } (\omega )$ and $G_{FF} (\omega )$ are non-negative for zero and non-zero frequency; this can be seen in the low temperature analytic solutions (see section \ref{LowEn}), as well as in numerics for various values of $\beta J$. Both terms mean that the solution is stable.

\section{Large $p$ analysis }

In this appendix,  we study the large $p$ limit of the $\mathcal{N} =2$ and $\mathcal{N} =4$ theories. We first discuss an ansatz for the large $p$ behavior which will work at high temperatures. This first ansatz fails at a certain critical temperature and will be replaced by another one that we discuss later. 

The first ansatz for the behavior of $G_{\phi } $ at large $p$ is
\begin{equation} \la{Gpan}
JG_{\phi } =e^{g_{\phi } /p} 
\end{equation}
where $g_{\phi } $ is finite, and  
\begin{equation} \label{eq:large_q_J_def}
J=\frac{2\cJ}{p} 
\end{equation}
with ${\cal J}$ finite in the large $p$ limit. 
Note that we have to restore the $J$ dependence, which was previously set to $J=1$ in some formulas, so that we can make this substitution. 

By \eqref{eq:SD_eqs_G_general_q} we have that
\begin{equation}
\Sigma _F(\tau )=-\frac{2\cJ}{p} e^{g_{\phi } } (\tau )
\end{equation}
where we always work up to order $1/p$. Then using \eqref{eq:SD_eqs_Sigma}
\begin{equation} \label{eq:large_q_eq_for_f}
G_F(\omega ) =1-\frac{2\cJ}{p} e^{g_{\phi } } (\omega )
\end{equation}
where we mean that $e^{g_{\phi } (\tau )} $ is evaluated in frequency space.

Moving on to the fermion, from \eqref{eq:SD_eqs_G_general_q}
\begin{equation}
\Sigma _{\psi } =\frac{4\cJ^2}{p} e^{g_{\phi } (\tau )} G_{\psi} (\tau ) .
\end{equation}
Then using \eqref{eq:SD_eqs_Sigma} (remembering that the frequencies are half integral)
\begin{equation}
G_{\psi} (\omega )=\frac{1}{-i\omega } -\frac{1}{\omega ^2} \frac{4\cJ^2}{p} \left( e^{g_{\phi } } \times G_{\psi} \right) (\omega )
\end{equation}
or in the time domain
\begin{equation} \label{eq:large_q_eq_for_psi}
G_{\psi} ''(\tau )=\delta '(\tau )+\frac{4\cJ^2}{p} e^{g_{\phi   } (\tau )} G_{\psi} (\tau ).
\end{equation}
Therefore, at leading order 
\be 
\la{Gpsian} 
G_{\psi} (\tau )=\frac{\sgn(\tau )}{2} +O(1/p) ~.
\ee 
Finally, let us consider the boson, for which using again \eqref{eq:SD_eqs_G_general_q}
\begin{equation}
\Sigma _{\phi } (\tau )=-\frac{4\cJ^2}{p} e^{g_{\phi } } \delta (\tau )+\frac{2\cJ^3}{p} e^{g_{\phi } }  .
\end{equation}
Now we plug this back in \eqref{eq:SD_eqs_Sigma}. We distinguish between the zero mode and the non-zero mode; for the zero mode
\begin{equation}
JG_{\phi } (\omega=0) = \frac{1}{2\cJ e^{g_{\phi } (\tau =0)} -\cJ^2 \int d\tau\,  e^{g_{\phi } (\tau )} } =\int d\tau \left( 1+\frac{g_{\phi } }{p} \right) =\beta +O(1/p) .
\end{equation}
This gives us an equation that determines the zero mode
\begin{equation} \label{eq:large_q_zero_mode_eq}
e^{g_{\phi } (\tau =0)} -\frac{\cJ}{2} \int_0^{\beta }  d\tau \, e^{g_{\phi } (\tau )} =\frac{1}{2\cJ \beta } .
\end{equation}
For the non-zero mode, we get, similarly to before, the equation
\begin{equation} \label{eq:large_q_Gphi_ODE}
\partial _{\tau } ^4 (J G_{\phi } )=-\frac{2\cJ}{p} \delta ''(\tau ) -\frac{8\cJ^3}{p^2} e^{g_{\phi } (\tau=0)} \delta (\tau )+\frac{4\cJ^4}{p^2} e^{g_{\phi } } .
\end{equation}
The last two terms are subleading. Without them, the boundary conditions $\partial _{\tau } G_{\phi } (0^+)=- \frac{1}{2} $ and $G_{\phi } (\beta -\tau )=G_{\phi } (\tau )$ fix the result. The latter condition means that $G_{\phi } $ is quadratic, and together with  the other boundary condition, at leading order we get
\begin{equation}
g_{\phi } =\text{const}-\cJ \tau +\frac{\cJ}{\beta } \tau ^2
\end{equation}
which is the free theory result. Eq.\ \eqref{eq:large_q_zero_mode_eq} determines the constant, so that for $0\le\tau \le\beta $
\begin{equation} \la{Glargep}
JG_{\phi } =1+\frac{g_{\phi } }{p} =1+\frac{1}{p} \Bigg\{ -\log \left[ 2\beta \cJ \left( 1-\frac{\sqrt{\pi \beta \cJ}}{2} e^{-\frac{\beta \cJ}{4} } \erfi\Big( \frac{\sqrt{\beta \cJ}}{2} \Big) \right) \right] -\cJ \tau +\frac{\cJ\tau ^2}{\beta } \Bigg\}.
\end{equation}
$\erfi$ is the imaginary error function defined by $\erfi(x) = \frac{2}{\sqrt{\pi }} \int _0^x dt \, e^{t^2}$.
For $\beta \cJ \ll 1$ we have $J G_{\phi } \approx 1-\frac{1}{p} \log(2\beta \cJ)=1-\frac{1}{p} \log(p\beta J)$ which agrees with \eqref{ValSe} (note that when expressed in terms of $J$, in the large $p$ limit it requires temperatures greater than $p$).

Note that at this order, except for the zero mode, we essentially get the free result for the boson. We can use \eqref{eq:large_q_Gphi_ODE} and get the $1/p^2$ term, but we will not do it here.

To complete the results for the correlations functions, we plug $g_{\phi } $ back into \eqref{eq:large_q_eq_for_f} and get
\begin{equation}
G_F = \delta (\tau ) - \frac{1}{\beta p} \left[ 1-\frac{\sqrt{\pi \beta \cJ}}{2} e^{-\frac{\beta \cJ}{4} } \erfi \Big( \frac{\sqrt{\beta \cJ}}{2} \Big) \right] ^{-1} e^{-\cJ \tau +\frac{\cJ \tau ^2}{\beta } } .
\end{equation}
Similarly, \eqref{eq:large_q_eq_for_psi} gives us $G_{\psi} $ when using the boundary conditions $G_{\psi} (0^+)=\frac{1}{2} $ and $G_{\psi} (\beta -\tau )=G_{\psi} (\tau )$ (we work in the range $0 \le \tau  \le \beta $)
\begin{equation}
G_{\psi} =\frac{1}{2} +\frac{1}{2p} -\frac{1 }{2p} \, \frac{e^{-\cJ \tau + \frac{\cJ \tau ^2}{\beta } } -\frac{\sqrt{\pi \beta \cJ}}{2} \left( 1-\frac{2\tau }{\beta } \right) e^{-\frac{\beta \cJ}{4} } \erfi \Big(\frac{\sqrt{\beta \cJ}}{2} \Big(1-\frac{2\tau }{\beta } \Big)\Big)}{1-\frac{\sqrt{\pi \beta \cJ}}{2} e^{-\frac{\beta \cJ}{4} } \erfi \Big(\frac{\sqrt{\beta \cJ}}{2} \Big) } .
\end{equation}

Interestingly, the zero mode of $g_\phi$ diverges as we approach a critical value of   $\beta \cJ$ around $3.41613$, where the argument of the logarithm in \nref{Glargep} becomes zero.  There is no real solution for higher values. 
We do not interpret this as a physical divergence, but just as meaning that the ansatz for the $p$ dependence that we have made in \nref{Gpan} \nref{eq:large_q_eq_for_f} \nref{Gpsian} becomes invalid near this region. As we approach this critical temperature,  we should make a different ansatz.

  At lower temperatures we can make the following ansatz for the $p$ dependence, which was inspired by the low temperature expansion of section \ref{LowEn}.  

The low temperature ansatz is
\begin{equation}
\begin{split}
JG_{\phi }(\tau ) = 1+ \frac{\log p+\mathfrak{g}_{\phi }(\tau ) }{p} + \cdots 
\end{split}
\end{equation}
 and all the rest $G_{\psi} $, $G_{F} $, $\Sigma _{\phi } $, $\Sigma _{\psi} $, and $\Sigma _{F} $ being finite, except for the zero mode of $\Sigma _{\phi} $ which goes as $1/p$. This means that at leading order
 \begin{equation}
 (JG_{\phi }(\tau ) )^p = p e^{\mathfrak{g}_{\phi }(\tau ) } .
 \end{equation}
 We keep the same definition of $\cJ$ \nref{eq:large_q_J_def}.
 
Equation \eqref{eq:SD_eqs_G_general_q} tells us that (we always work at leading order)
 \begin{equation}
 \Sigma _F(\tau ) = -2 \cJ e^{\mathfrak{g}_{\phi } (\tau )} 
 \end{equation}
 or
 \begin{equation}
 G_F(\omega ) = \frac{1}{1+2 \cJ (e^{\mathfrak{g}_{\phi } }) (\omega )} 
 \end{equation}
 which is indeed $O(1)$.
 
 For the fermion,
 \begin{equation}
\Sigma _{\psi} (\tau ) = 4\cJ^2 G_{\psi} (\tau ) e^{\mathfrak{g}_{\phi } (\tau )} 
 \end{equation}
 and
 \begin{equation}
 G_{\psi} (\omega ) = \frac{1}{-i\omega -4\cJ^2 (G_{\psi} \times e^{\mathfrak{g}_{\phi } } )(\omega )} .
 \end{equation}
 
 Finally, for the boson
 \begin{equation}
 \Sigma _{\phi } (\tau ) = -4\cJ^2 G_F e^{\mathfrak{g}_{\phi } } +8\cJ^3 G_{\psi} ^2 e^{\mathfrak{g}_{\phi } }.
 \end{equation}
 Note that at leading order for the zero mode, the two terms above should cancel; $G_{\phi } $ should be $\frac{p}{2\cJ} +\frac{\log p+\mathfrak{g}_{\phi } (\tau )}{2\cJ} $. For the non-zero mode, we have the equation
 \begin{equation}
\frac{\mathfrak{g}_{\phi } (\omega )}{2\cJ} = \frac{1}{\omega ^2+4\cJ^2 (G_F \times e^{\mathfrak{g}_{\phi } } )(\omega )-8\cJ^3 (G_{\psi} \times G_{\psi} \times e^{\mathfrak{g}_{\phi } })(\omega ) } .
 \end{equation}
After expressing $G_{\psi } $ in terms of $\mathfrak{g}_{\phi}$, this is an equation for $\mathfrak{g}_{\phi } $ that is independent of $p$.

 We can see that the large $p$ limit of the   low temperature result \nref{FinG} means that in the   large $\beta {\cal J}$  limit we get 
\be 
\mathfrak{g}_\phi = 2 \log { \beta {\cal J }   \over 6 \pi }  + \cdots 
\ee 
where the dots are higher order terms. For a more complete expression see \nref{GpLaA}.

   \section{Comment on more general potentials } 
   \la{GenPotA} 
   
   Here we discuss the low energy expansion for more general potentials. Namely, we can consider a model which has random tensors of various orders, with various numbers of indices, or various values of $p$. 
This would lead to a term of the following form in the effective action 
   \be \la{GenPot}
 \beta \int d\tau   G_F ( d_2 G_\phi + \tilde d_3   G_\phi^2 +  \tilde d_4  G_\phi^3 + \cdots ) 
 \ee 
If we further expand each term using \nref{Gexpn} then we could get $\alpha$, $d_3$ and $d_4$  in terms of the $\tilde d_p$ coefficients, where the $d_p$ are defined through the terms \nref{QuadTe}, \nref{CubTe} and \nref{Quartic}.  
And we can then write the final answer as 
    \be \la{Expr}
 {\log Z \over N } = \half \log(p-1) + { \pi \over 2 \beta \sqrt{\alpha }  } + { d_3 \bG \beta^2 \over 12 \alpha }  - { d_3^2 \over 3 \alpha} - {   d_ 4 \over 4 \alpha^2  } 
 \ee

  \section{Higher order corrections to the low temperature solution} 
  \la{HigherOrder}

By solving the large $N$ equations \eqref{eq:SD_eqs_Sigma} and \eqref{eq:SD_eqs_G_general_q} at low temperature, we found the form of the correlation functions as an expansion in large $\beta J$, similarly to \cite{Lin:2013jra}. Here we will write the first few terms from what we obtained.  
The correlation function of the boson is
\begin{equation} \label{eq:low_T_time_domain_phi}
\begin{split}
& JG_{\phi } (\tau ) =
J^{\frac{2}{p+2} } \bG  \\
& - \frac{1 }{2\pi \sqrt{p-1}\bG^{\frac{p-2}{p} }} J^{-\frac{p-2}{p+2} } \log\left( 4\sin^2 \frac{\pi \tau }{\beta } \right)  
-J^{-\frac{2p-2}{p+2} } \frac{p-2}{16\pi ^2(p-1)\bG^{p-1} } \left[ \left( \log(4\sin^2 \frac{\pi \tau }{\beta } )\right) ^2-\frac{\pi ^2}{3} \right] -\\
&- J^{-\frac{2p-2}{p+2} } \frac{p}{8\pi ^2(p-1)\bG^{p-1} } \left( \frac{\pi ^2}{3} -\frac{2\pi ^2\tau }{\beta } +\frac{2\pi ^2\tau ^2}{\beta ^2} \right) + \cdots .
\end{split}
\end{equation}
The first term is the zero mode.
For the fermion
\begin{equation} \label{eq:low_T_time_domain_psi}
\begin{split}
G_{\psi} (\tau ) = J^{-\frac{2p}{p+2} } \frac{1}{\sqrt{p-1}\bG^{\frac{p-2}{2} } }  \frac{1}{\beta \sin \frac{\pi \tau }{\beta } } +J^{-\frac{3p}{p+2} } \frac{p-2}{4\pi\beta (p-1)\bG^{p-1}  } \frac{\log \left( 4 \sin ^2 \frac{\pi \tau }{\beta } \right) }{\sin \frac{\pi \tau }{\beta } } +\cdots .
\end{split}
\end{equation}
For the auxiliary field
\begin{equation} \label{eq:low_T_time_domain_f}
\begin{split}
& J^{-1} G_F(\tau ) = J^{-\frac{4p+2}{p+2} } \frac{1}{\beta^2 \bG^{p-1}  } -J^{-\frac{3p+2}{p+2} } \frac{\pi }{\sqrt{p-1} \bG^{\frac{p-2}{2} } }  \frac{1}{\beta ^2\sin^2 \frac{\pi \tau }{\beta } } -\\
& - J^{-\frac{4p+2}{p+2} } \frac{p-2}{2\beta^2 (p-1)\bG^{p-1}  } \left( 1-\frac{1}{\sin ^2 \frac{\pi \tau }{\beta } } +\frac{1}{2} \frac{\log\left( 4\sin ^2 \frac{\pi \tau }{\beta } \right) }{\sin ^2 \frac{\pi \tau }{\beta } } \right)  + J^{-\frac{4p+2}{p+2} } \frac{p}{2(p-1)\beta  \bG^{p-1} } \left( \delta (\tau )-\frac{1}{\beta } \right) +\cdots .
\end{split}
\end{equation}

Similarly, the self energies are given by
\begin{equation}
\begin{split}
&J^{-3} \Sigma _{\phi } =-J^{-\frac{2p+6}{p+2} } \frac{1}{\beta^2\bG } +J^{-\frac{p+6}{p+2} } \frac{\pi \sqrt{p-1}\bG^{\frac{p-2}{2} } }{\beta ^2 \sin ^2 \frac{\pi \tau }{\beta } } -\\
& \qquad - J^{-\frac{2p+6}{p+2} } \frac{p-2}{2\beta^2 \bG } \left( 1-\frac{1}{\sin^2 \frac{\pi \tau }{\beta } } +\frac{1}{2} \frac{\log\left( 4 \sin^2 \frac{\pi \tau }{\beta }\right)  }{\sin^2 \frac{\pi \tau }{\beta } } \right)  - J^{-\frac{2p+6}{p+2} } \frac{p}{2\beta \bG} \left( \delta (\tau )-\frac{1}{\beta } \right) +\cdots ,\\
& J^{-2} \Sigma _{\psi} = J^{-\frac{4}{p+2} } \frac{\sqrt{p-1}\bG^{\frac{p-2}{2} } }{\beta \sin \frac{\pi \tau }{\beta } } -J^{-\frac{p+4}{p+2} } \frac{p-2}{4\pi \beta \bG} \frac{\log\left( 4\sin^2 \frac{\pi \tau }{\beta } \right) }{\sin \frac{\pi \tau }{\beta } } +\cdots ,\\
& J^{-1} \Sigma _{F} = -J^{\frac{2p-2}{p+2} }\bG^{p-1} -J^{-\frac{2}{p+2} } \frac{p-2}{24\bG} +J^{\frac{p-2}{p+2} } \frac{\sqrt{p-1}\bG^{\frac{p-2}{2} } }{2\pi } \log\left( 4\sin^2 \frac{\pi \tau }{\beta } \right)  - \\
& - J^{-\frac{2}{p+2} } \frac{p-2}{8\pi ^2\bG} \left[ \frac{1}{2} \left( \log (4\sin^2 \frac{\pi \tau }{\beta } )\right) ^2-\frac{\pi ^2}{6} \right]  + J^{-\frac{2}{p+2} } \frac{p}{8\pi ^2\bG} \left( \frac{\pi ^2}{3} -\frac{2\pi ^2\tau }{\beta } +\frac{2\pi ^2\tau ^2}{\beta ^2} \right) +\cdots .
\end{split}
\end{equation}

We now observe that with \eqref{eq:large_q_J_def}, we have a well defined large $p$ limit for these results, giving
\bea \la{GpLaA}
&~&  JG_{\phi } (\tau ) = 1+\frac{\log p + \mathfrak{g}_\phi }{p} \cr 
&~& \mathfrak{g}_\phi =2\log \frac{\beta \cJ}{6\pi }-\frac{3}{ \beta \cJ } \log\left( 4\sin^2 \frac{\pi \tau }{\beta } \right)  - \frac{9}{4 (\beta \cJ)^2} \left[ \left( \log(4\sin^2 \frac{\pi \tau }{\beta } )\right) ^2+\frac{\pi ^2}{3} -\frac{4\pi ^2\tau }{\beta } +\frac{4\pi ^2\tau ^2}{\beta ^2} \right] + \cdots ~~~~ ~~~~~\\
&~&  G_{\psi} (\tau ) = \frac{3\pi }{(\beta \cJ)^2 \sin \frac{\pi \tau }{\beta } } +\frac{9\pi }{2 (\beta \cJ)^3} \frac{\log \left( 4 \sin ^2 \frac{\pi \tau }{\beta } \right) }{\sin \frac{\pi \tau }{\beta } } +\cdots ,\cr 
&~&  \cJ^{-1} G_F(\tau )= \frac{18\pi ^2}{(\beta\cJ) ^4} -\frac{3\pi ^2}{(\beta \cJ)^3} \frac{1}{\sin^2 \frac{\pi \tau }{\beta } } - \frac{9\pi ^2}{(\beta \cJ)^4} \left( -\beta \delta (\tau ) +2-\frac{1}{\sin ^2 \frac{\pi \tau }{\beta } } +\frac{1}{2} \frac{\log\left( 4 \sin ^2 \frac{\pi \tau }{\beta } \right) }{\sin ^2 \frac{\pi \tau }{\beta } } \right) +\cdots .
\eea
The self energies are  
\begin{equation}
\begin{split}
&\cJ^{-3} \Sigma _{\phi } =-\frac{2}{p(\beta {\cal J})^2 } +\\
& \qquad \qquad +  \frac{1 }{3\beta \cJ \sin ^2 \frac{\pi \tau }{\beta } } - \frac{1}{(\beta \cJ)^2 } \left( 1-\frac{1}{\sin^2 \frac{\pi \tau }{\beta } } +\frac{1}{2} \frac{\log\left( 4 \sin^2 \frac{\pi \tau }{\beta }\right)  }{\sin^2 \frac{\pi \tau }{\beta } } \right)  - \frac{1}{\beta \cJ^2} \left( \delta (\tau )-\frac{1}{\beta } \right) +\cdots ,\\
& \cJ^{-2} \Sigma _{\psi} =  \frac{1}{3\pi \sin \frac{\pi \tau }{\beta } } -\frac{1}{2\pi \beta \cJ } \frac{\log\left( 4\sin^2 \frac{\pi \tau }{\beta } \right) }{\sin \frac{\pi \tau }{\beta } } +\cdots ,\\
& \cJ^{-1} \Sigma _{F} = - \frac{(\beta \cJ)^2}{18\pi ^2} - \frac{1}{12} +\frac{\beta \cJ}{6\pi^2} \log\left( 4\sin^2 \frac{\pi \tau }{\beta } \right)  - \\
& \qquad \qquad  - \frac{1}{4\pi ^2} \left[ \frac{1}{2} \left( \log (4\sin^2 \frac{\pi \tau }{\beta } )\right) ^2-\frac{\pi ^2}{6} \right]  + \frac{1}{4\pi ^2} \left( \frac{\pi ^2}{3} -\frac{2\pi ^2\tau }{\beta } +\frac{2\pi ^2\tau ^2}{\beta ^2} \right) +\cdots .
\end{split}
\end{equation}

    \section{Full equations involving replicas } 
    
      In this appendix,  we write all the equations for the various configurations we discussed in section \ref{Replicas}. 
    
    \subsection{Full equations for the replica symmetric case } 
    \la{FullRSB}
    
    The equations for $G_{\psi } $ and $G'_F$ are the same as in the analysis without replicas, except that $ G_F'(\omega =0) =0$,   while the equation for $G_{\phi } $ is modified. We will not need this equation, but we write it for completeness  
\begin{equation}
\begin{split}
& G_{\phi } (\omega )^{-1} =\omega ^2 +  \left[ (p-1)(G_{\phi } ^{p-2} G'_F)(\omega )-(p-1)(p-2)(G_{\psi } ^2 G_{\phi } ^{p-3} )(\omega )\right] + \\
& \qquad + \frac{(p-1)  \beta ^{-1} (G_{\phi } ^{p-2} )(\omega )}{1+  M-  \beta \bar G_{\phi }^{p-1} x^{p-1} } 
 - (p-1)  \bar G_{\phi } ^{p-1} x^{p-1} \frac{(G_{\phi } ^{p-2} )(\omega )-\beta \bar G_{\phi }^{p-2} x^{p-1} \delta _{\omega ,0} }{\left( 1+  M-  \beta \bar G_{\phi }^{p-1} x^{p-1} \right)^2} .
\end{split}
\end{equation}
By $(G_{\phi } ^{p-2} )(\omega )$ we mean the function $G_{\phi } (\tau )^{p-2} $ evaluated at frequency $\omega $, and similarly for the other functions.

    \subsection{Full equations for the 1 step replica symmetry breaking case } 
    
   \la{Full1RSB}

   Let us define for simplicity
\begin{equation}
\begin{split}
& A = 1+M- Y x^{p-1} ~,~~~~~~~~Y = \beta \bG^{p-1} \\
& B = 1+M+(m-1) Y x^{p-1} -m Y y^{p-1}.
\end{split}
\end{equation}

We have given the $y$ and $x$ equations in \nref{eqny} and \nref{eqnx}. 
%
The $m$ equation is
\begin{equation} \label{eq:1RSB_m_eq}
\begin{split}
& \frac{1}{2m^2} \log(1-x)-\frac{1}{2m^2} \log\left( 1+(m-1)x-my\right)  - \frac{1}{2m^2} \log \frac{A}{B} +\frac{1}{2m} \frac{x-y}{1+(m-1)x-my} -\\
& - \frac{1}{2} \frac{y(x-y)}{\left( 1+(m-1)x-my\right) ^2} -\frac{1}{2m} \frac{Y \left( x^{p-1} -y^{p-1} \right) }{B} 
+ \frac{1}{2} \frac{  Y^2 y^{p-1} \left( x^{p-1} -y^{p-1} \right) }{B^2} = 0.
\end{split}
\end{equation}
The equation of motion for $G_\phi $ is
\begin{equation}
\begin{split}
& G_{\phi } (\omega )^{-1} =\omega ^2 + \left[ (p-1)(G'_F G_{\phi } ^{p-2} )(\omega ) - (p-1)(p-2)(G_{\psi } ^2G_{\phi } ^{p-3} )(\omega )\right]  \\
&+\frac{m-1}{m} \frac{(p-1) \left[ \beta ^{-1} (G_{\phi } ^{p-2} )(\omega ) - \delta _{\omega ,0} \bG^{p-2} x^{p-1} \right] }{A} \\
& + \frac{1}{m} \frac{(p-1)   \left[ \beta ^{-1} (G_{\phi } ^{p-2}) (\omega )+\delta _{\omega ,0} (m-1) \bG^{p-2} x^{p-1} \right] }{B} \\
& - \frac{1}{B^2} (p-1)Y  y^{p-1} \left[ \beta ^{-1} (G_{\phi } ^{p-2} )(\omega )+\delta _{\omega ,0}\bG^{p-2} \left( (m-1)x^{p-1} -my^{p-1} \right) \right]  .
\end{split}
\end{equation}
The equations for $G_\psi ,G'_F$ are the same as previously, and we remember that   $G'_F(\omega=0)=0$.

Again, as a consistency check, when $x=y$, nothing depends on $m$, the equations for $x$ and $y$ become the same as the equation for $x$ in the replica symmetric case, and similarly for the equation for $G_{\phi } $. When $m=1$, the equation for $y$ reduces to the equation for $x$ in the replica symmetric case, and similarly for $G_{\phi } $.

\bibliographystyle{apsrev4-1long}
\bibliography{GeneralBibliography.bib}
\end{document}